\documentstyle[12pt,aaspp4,amsfonts]{article} 

\lefthead{\sc de Grijs, O'Connell \& Gallagher}
\righthead{\sc The Fossil Starburst in M82}

\begin{document}

\def\asec {{$\buildrel{\prime\prime}\over .$}}

\title{The Fossil Starburst in M82\footnote{Based on observations with
the NASA/ESA {\sl Hubble Space Telescope}, obtained at the Space
Telescope Science Institute, which is operated by the Association of
Universities for Research in Astronomy (AURA), Inc., under NASA contract
NAS 5-26555.}}

\author{Richard de Grijs\footnote{Present address: Institute of
Astronomy, University of Cambridge, Madingley Road, Cambridge CB3 0HA,
UK}, Robert W.  O'Connell} \affil{Astronomy Department, University of
Virginia, P.O.\ Box 3818, Charlottesville, VA 22903-0818;
rd7a@astsun.astro.virginia.edu, rwo@virginia.edu}
\and
\author{John S. Gallagher, {\sc iii}}
\affil{Astronomy Department, University of Wisconsin, 475 North Charter
Street, Madison, WI 53706; jsg@astro.wisc.edu}

\begin{abstract} 

We present high-resolution {\sl HST} imaging in the optical ({\sl
WFPC2}) and near-infrared ({\sl NICMOS}) of a disk region 1 kpc NE of
the starburst core in the nearby galaxy M82.  This region, M82 ``B,''
has been suspected to be a fossil starburst site in which an intense
episode of star formation occurred over 100 Myr ago, and our new
observations confirm this intepretation.  M82 thus presents us with the
opportunity to observe both active and evolved starburst environments at
close range. 

The surface brightness of M82 B is well above normal for galactic disks
and comparable to the core surface brightnesses in spiral galaxies.  Its
intrinsic surface brightness at an age of 10 Myr was comparable to that
found in the present-day nuclear starburst, indicating an event of
comparable amplitude. 

We find a large, evolved system of super star clusters in M82 B.  Using
size as a criterion to distinguish cluster candidates from point
sources, we identify a total of 113 super star cluster candidates.  We
use a two-color {\it BVI} diagram and evolutionary spectral synthesis
models to separately estimate the extinction and age of each cluster. 
The clusters range in absolute magnitude from $M_V^0 = -6$ to --10, with
a peak at --7.5.  The derived age distribution suggests steady,
continuing cluster formation at a modest rate at early times ($> 2$ Gyr
ago), followed by a concentrated formation episode $\sim 600$ Myr ago
and more recent suppression of cluster formation.  The peak episode
coincides with independent dynamical estimates for the last tidal
encounter with M81, which presumably induced the starburst. 

Our {\it J} and {\it H} band observations resolve the bright giant
population in M82's disk for the first time.  Star formation evidently
continued in M82 B until about 20 -- 30 Myr ago, but none is found
associated with the youngest generations in the nuclear starburst (age
$\lesssim 15$ Myr). 

After correcting the cluster luminosity function to a fiducial age of 50
Myr, we find that the bright end is characterized by a power-law slope
with $\alpha = -1.2 \pm 0.3$, similar to that of other young cluster
systems in interacting galaxies.  There is tentative evidence for
broadening of the luminosity function due to dynamical destruction of
lower mass clusters.  Cluster sizes ($2.34 \le R_{\rm eff} \lesssim 10$
pc, or $2.4 \lesssim R_{\rm core} \lesssim 7.9$ pc) and estimated masses
(a median of $10^5 M_\odot$) are consistent with values found for young
super star cluster populations in M82's core and other galaxies and with
the progenitors of globular clusters. 

\end{abstract}


\section{Introduction: Multiple Starbursts in M82}

\subsection{A fossil starburst?}
\label{intro1.sect}

M82 is the prototype starburst galaxy.  Observations at all wavelengths
from radio to X-rays (reviewed in Telesco 1988 and Rieke et al.\ 1993)
are consistent with the following scenario.  During the last several 100
Myr, tidal interactions between M82 and M81 and/or other member galaxies
of the M81 group have channeled large amounts of gas into the central
regions of M82.  This has induced a starburst which has continued for up
to about 50 Myr at a star formation rate of $\sim 10$ M$_\odot$
yr$^{-1}$.  Energy deposited by supernovae, at a rate of $\sim 0.1$
supernova yr$^{-1}$ (e.g.\ O'Connell \& Mangano 1978, hereinafter OM;
Rieke et al.  1980, McLeod et al.\ 1993), drives a large-scale galactic
wind along the minor axis of M82 (e.g.\ Lynds \& Sandage 1963, Fabbiano
\& Trinchieri 1984, Watson, Stanger, \& Griffiths 1984, McCarthy,
Heckman, \& van Breugel 1987, Shopbell \& Bland-Hawthorn 1998,
Strickland, Ponman, \& Stevens 1997).  All of the bright radio and
infrared (IR) sources associated with the active starburst are confined
to the galaxy's center.  They lie within a radius of $\sim 250$ pc and
mostly suffer heavy extinction by dust, estimated to be in the range
${\rm A}_{\rm V} \sim$ 5--25 (e.g.  Telesco et al.\ 1991, McLeod et al.\
1993, Satyapal et al.\ 1995).  They correspond spatially with bright
optical structures, labeled M82 A, C, and E in OM.  These are probably
those parts of the starburst core which happen to be the least obscured
along the line of sight (O'Connell et al.\ 1995). 

However, there is good photometric and spectroscopic evidence that this
was not the only major starburst episode to have occurred in M82.  A
region about 1 kpc NE from the galactic center, M82 B (cf.\ OM and Fig. 
1), has exactly the properties one would predict for a fossil starburst
with an amplitude similar to the active burst in M82 A, C and E.  M82 B
has an intrinsic surface brightness of $\mu_V^0 \sim 16.5$ mag
arcsec$^{-2}$ (measured in a $7''$ aperture).  Although this is fainter
than the intrinsic {\it V}-band surface brightness of M82 A, C, and E in
the active starburst, it is nonetheless much higher than the maximum
brightnesses encountered in spiral disks, even when seen edge-on.  For
the disks of typical edge-on Sb-Sd galaxies, de Grijs (1998) finds
central surface brightnesses of order $\mu_B \sim 20 - 22$ and $\mu_I
\sim 18 - 20$ mag arcsec$^{-2}$.  The surface brightness of M82 B is, in
fact, comparable to the {\it nuclear} brightnesses of bright Sb-Sc
galaxies measured with similar resolution (e.g.\ de Jong 1996).  By
extrapolating region B's surface brightness back to an age of 10 Myr,
using the evolutionary synthesis models of Bruzual \& Charlot (1996), we
estimate that its surface brightness was of order $\mu_V^0 \sim 14.5$
mag arcsec$^{-2}$, similar to that presently observed in the active
starburst M82 A. 

The spectrum of M82 B is likewise consistent with a fossil starburst
event, entirely comparable to that now transpiring in M82 A.  It is
dominated by Balmer absorption lines and shows a large Balmer
discontinuity (OM, Marcum \& O'Connell 1996).  Emission lines are very
weak, unlike in M82 A, which has intense line emission.  These are the
hallmarks of the anomalous ``E+A'' spectra found in distant galaxy
clusters that exhibit the Butcher-Oemler effect (Butcher \& Oemler 1978,
Oemler 1992).  Such spectra are generally interpreted as the signature
of a truncated burst of star formation that occurred 100--1000 Myr
earlier (e.g.\ Couch \& Sharples 1987, Dressler \& Gunn 1990).  These
starbursts are thought to be part of the process by which disk galaxies
are converted to elliptical or lenticular galaxies and probably result
from tidal interactions, mergers, or perhaps ram-pressure stripping by
the intergalactic medium (Butcher \& Oemler 1978, Oemler 1992, Barger et
al.\ 1996).  From 20--40 \AA\ resolution spectrophotometry, OM and
Marcum \& O'Connell (1996) find evidence of a sharp main-sequence
truncation in M82 B, corresponding to a turnoff age of $\sim$ 100--200
Myr and an average extinction of $A_V \sim 0.6$ mag.  Region A's
spectrum, on the other hand, indicates a very young population ($\sim 5$
Myr) and is more heavily affected by extinction ($A_V \sim 2.2$ mag). 

Apart from the physical properties of M82 B, three other considerations
support the notion that multiple starbursts have occurred in M82. 
First, the triggering gravitational interactions between M82 and other
members of the M81 group and the associated tidal gas flows have been
proceeding for more than 200 Myr (e.g.\ Yun, Ho, \& Lo 1994), in which
time multiple starbursts could easily have occurred.  Second, starbursts
are unlikely to be synchronized throughout the disturbed galaxy, and
Satyapal et al.\ (1997) find evidence of outward propagation of star
formation in the dense core of M82.  Finally, it is likely that
starbursts are strongly self-limited, or quenched, by supernova-driven
outflows, which remove the remaining cool gas from the immediate
starburst region (e.g.\ Chevalier \& Clegg 1985, Doane \& Mathews 1993). 
The remarkable minor-axis wind in M82 is a dramatic example of this
process.  However, the disturbed conditions near an early burst may
discourage re-ignition at the same site when cool gas inflows resume,
shifting the location of active star formation. 

Although other interesting examples of relatively nearby post-starburst
systems have been located (e.g.\ Vigroux, Boulade, \& Rose 1989,
Oegerle, Hill, \& Hoessel 1991, Caldwell et al.\ 1993, Caldwell \& Rose
1997) none of these, nor any other nearby starburst, offer the
opportunity to study two discrete starbursts at such close range as M82. 

The active starburst in the inner 250 pc of M82 has been studied
thoroughly, whereas the fossil starburst in region B has not received
much attention.  The aim of this paper is to study the remains of this
earlier episode in detail, with emphasis on young star cluster
candidates. 

\subsection{Super Star Clusters in M82}

Using {\sl Hubble Space Telescope (HST)} imaging of the bright central
regions A, C, and E of M82, O'Connell et al.\ (1995) resolved these into
a swarm of over 100 young star cluster candidates, with a mean $L_V \sim
4 \times 10^6 L_\odot$, brighter than any globular cluster in the Local
Group.  Satyapal et al.\ (1997) found evidence in the near-IR of a dozen
compact clusters lying deeper within the heavily obscured core of the
galaxy near A, C, and E.  The core clusters show an age dispersion of 6
Myr and indications of outward propagation of star formation.  Gallagher
\& Smith (1999) determined a significantly older age of 60 Myr for the
luminous cluster F, which is located 440 pc South-West of the nucleus. 

Such ``super'' star clusters (SSCs) have been discovered, mostly with
{\sl HST}, in other interacting, amorphous, dwarf, and starburst
galaxies (e.g.\ Arp \& Sandage 1985, Melnick, Moles, \& Terlevich 1985,
Meurer et al.\ 1992, Holtzman et al.\ 1992, Whitmore et al.\ 1993, 1999,
Hunter et al.\ 1994, 2000, O'Connell, Gallagher, \& Hunter 1994, Ho \&
Filippenko 1996b, Conti, Leitherer, \& Vacca 1996, Ho 1997, Watson et
al.\ 1996, Carlson et al.\ 1998).  Their diameters, luminosities, and --
in several cases -- masses are consistent with these being young {\it
globular} clusters formed as a result of recent gas flows (e.g.\ van den
Bergh 1995, Meurer 1995, Ho \& Filippenko 1996a, Hunter et al.\ 2000). 
In the case of M82 F, Smith \& Gallagher (2000) obtained a dynamical
mass of $2 \times 10^6 M_\odot$.  It is possible that a large fraction
of the star formation in starbursts takes place in the form of such
concentrated clusters.  The discovery that globular cluster formation,
once thought to occur only during early stages of galaxy evolution,
continues today is one of {\sl HST}'s main contributions to astrophysics
to date. 

M82 is one of the few galaxies containing a rich, starburst-induced
cluster system which is near enough that the internal structure of
individual SSCs can be studied.  Most cluster systems are too distant
for such work, even with {\sl HST}.  Apart from their intrinsic
interest, SSCs can be individually age-dated and are therefore important
tracers of the history of the starburst process across the face of a
galaxy like M82. 

If region B is indeed a fossil starburst site, it probably contains
evolved clusters which originally had properties similar to those now
observed in region A.  The presence of both the active and the fossil
starburst sites in M82 therefore provides a unique arena for the study
of the stellar and dynamical evolution of star cluster systems.  Even in
the time interval of 0.3--1 Gyr that has probably elapsed since the
starburst in region B, evolutionary effects might be detectable because
of the dense environment of the starburst.  In order to explore the
fossil starburst site and to detect candidate SSCs, we therefore made
new, multiband optical/near-IR observations of region B with {\sl HST}. 

In \S 2, we provide an overview of the observations and the image
processing techniques applied.  \S\S 3 and 4 describe the general
characteristics of region B; \S 5 gives an outline of the analysis
techniques used to select our star cluster candidates and obtain the
results presented in the subsequent sections.  We report the first
detection of the resolved field stellar population in the disk of M82 in
\S 6.  A discussion of the ages of the star cluster sample, in terms of
evolutionary spectral synthesis models is presented in \S 7.  The
cluster luminosity functions and estimated masses are discussed in \S\S
8 and 9, respectively.  In \S 10 we present a detailed analysis of the
cluster profiles.  Finally, in \S 11 we summarize our results and
conclusions. 

\section{Observations and Data Reduction}

\subsection{Broad-Band HST Imaging}

We observed M82 B on September 15, 1997, with both the {\sl Wide Field
Planetary Camera 2 (WFPC2)} and the {\sl Near-Infrared Camera and
Multi-Object Spectrometer (NICMOS)} on board the {\sl HST} (program ID
7446).  We imaged two adjacent 35\arcsec\ square fields, ``B1'' and
``B2'' with the Planetary Camera (PC) chip (0\asec0455 pix$^{-1}$) of
the {\sl WFPC2}.  Field centers (J2000) of the PC were RA = $09^{\rm
h}56^{\rm m}02.^{\rm s}695$, Dec = $+69^\circ41' 10$\asec05 for B1 and
RA = $09^{\rm h}55^{\rm m}57.^{\rm s}195$, Dec = $+69^\circ41'
06$\asec35 for B2.  We used the F439W, F555W and F814W passbands, with
total integration times of 4400s, 2500s and 2200s, respectively, for
region B1 and 4100s, 3100s and 2200s, respectively, for region B2.  The
field of view during the observations was rotated clockwise by,
respectively, 19.09$^{\circ}$ and 20.55$^{\circ}$ from the North-South
axis.  These observations were obtained with four exposures per filter,
using two exposure times, to facilitate the removal of cosmic ray events
and increase dynamic range.  The {\sl WFPC2} F439W, F555W and F814W
filters are roughly comparable to the Johnson-Cousins broad-band {\it B,
V} and {\it I} filters, respectively. 

In Fig.  \ref{ground.fig} we have overlaid the contours of our two {\sl
WFPC2} fields on a {\it B}-band image, extracted from a Palomar 5m plate
taken by Sandage (exposure time 20 minutes, seeing $\lesssim 1''$; cf.\
O'Connell et al.\ 1995).  The PC CCD chip is the square ``step'' in the
overall L-shaped footprint of the {\sl WFPC2}.  The contours are labeled
by the part of region B which is centered on the PC.  ``B1'' is an
irregular, high surface brightness region lying along the major axis of
the galaxy about 55\arcsec\ (960 pc) NE of the brightest optical region,
A.  [For the distance of M82, we adopt 3.6 Mpc, or $m-M = 27.8$ mag,
based on the Cepheid distance for M81 obtained by Freedman et al.\
(1994); see also Sakai \& Madore (1999).  The corresponding linear scale
is 17.5 pc per arcsec.] Region B2 is halfway between B1 and A and is
therefore closer to the nuclear starburst.  It has lower {\it B}-band
surface brightness than B1 and is more enveloped in complex dust lanes. 
The conspicuous central dust lane which bisects the galaxy marks the
western edge of B2. 

In the near-IR we chose to use {\sl NICMOS} Camera-2 (0\asec075
pix$^{-1}$) as the best compromise of resolution and field of view.  For
both B1 and B2, we acquired 4 partially overlapping exposures in both
the F110W and F160W filters (comparable to the Bessell {\it J} and {\it
H} filters, respectively), in a tiled pattern.  The integrations, with
effective integration times of 768s for each field and filter, were
taken in MULTIACCUM mode to preserve dynamic range and to correct for
cosmic rays.  Sky background exposures were not necessary because of the
brightness of the sources. 

\subsection{Image Processing}

Pipeline image reduction and calibration was done with standard
procedures by the Space Telescope Science Institute (STScI), using the
updated and corrected on-orbit flat fields most appropriate for our
observations. 

Since our {\sl WFPC2} images with a common pointing were aligned to
within a few hundredths of a pixel, we simply co-added the individual
observations in a given filter using the IRAF/STSDAS\footnote{The Image
Reduction and Analysis Facility (IRAF) is distributed by the National
Optical Astronomy Observatories, which is operated by the Association of
Universities for Research in Astronomy, Inc., under cooperative
agreement with the National Science Foundation.  STSDAS, the Space
Telescope Science Data Analysis System, contains tasks complementary to
the existing IRAF tasks.  We used Version 2.0 (September 1997) for the
data reduction performed in this paper.} task {\sc crrej}.  This task
also removed cosmic ray events in a series of iterations that allow
correction for cosmic ray hits in pixels adjacent to those that already
have been corrected in an earlier iteration.  After some experimenting,
we found that as few as 2--3 iterations produce output images in which
the remaining cosmic ray hits -- if any -- are indistinguishable from
poisson noise in the relatively bright background levels of our images. 

Due to the smaller field of view of {\sl NICMOS} Camera-2, we had to
mosaic the four tiled exposures at each position to cover an area
comparable to the PC field.  The mosaicing was done using objects in
common on the frames for the determination of the exact spatial offsets. 
The flux histogram in the overlapping area was used to determine the
necessary adjustment of background levels (see de Grijs 1997, 1998). 

The rebinning (under conservation of the observed flux) and rotation of
the {\sl NICMOS} images to the {\sl WFPC2} pixel size and orientation,
and the alignment of the optical and near-IR images were performed using
the standard IRAF tasks {\sc magnify, rotate}, and {\sc imalign}.  We
determined the rotation angles and alignment offsets using objects in
common on optical and near-IR images.  The final, aligned images were
trimmed to a standard size ($759 \times 747$ pixels, or 34\asec$53
\times 33$\asec99) to facilitate object detection in a common reference
frame. 

\subsection{Photometric Calibration}
\label{calib.sect}

We used the transformation coefficients of Holtzman et al.\ (1995) to
convert measurements made from the {\sl WFPC2} images to the standard
{\sl B,V, I} photometric system:
\begin{eqnarray}
\label{standB.eq}
B &=& -2.5 \times \log \dot{C}({\rm F439W}) +
(0.003 \pm 0.007) \times (B-V) \nonumber \\
&+& (-0.088 \pm 0.003) \times (B-V)^2 + (20.070 \pm 0.004) + 2.5 \times \log
({\rm GR}) \, ,
\end{eqnarray}
\begin{eqnarray}
V &=& -2.5 \times \log \dot{C}({\rm F555W}) +
(-0.052 \pm 0.007) \times (V-I) \nonumber \\
&+& (0.027 \pm 0.002) \times (V-I)^2 + (21.725 \pm 0.005)  + 2.5 \times \log
({\rm GR}) \, ,
\end{eqnarray}
and
\begin{eqnarray}
\label{standI.eq}
I &=& -2.5 \times \log \dot{C}({\rm F814W}) +
(-0.062 \pm 0.009) \times (V-I) \nonumber \\
&+& (0.025 \pm 0.002) \times (V-I)^2 + (20.839 \pm 0.006) + 2.5 \times \log
({\rm GR}) \, .
\end{eqnarray}
Here, $\dot{C}$ is the sum of the pixel values from the processed images
in the spatial area of interest divided by the integration time, and GR
is the gain ratio as defined by Holtzman et al.\ (1995).  For the {\sl
PC} chip, with an analog-to-digital gain of 7 electrons (as for our
observations) GR $= 1.987$.  These transformations hold for $-0.3 <
(B-V) < 1.5$ and $-0.3 < (V-I) < 1.5$.  The cluster colors in M82 B are
well within these colors ranges (\S \ref{colors.sect}).  The color terms
in Eqs.\ (\ref{standB.eq})--(\ref{standI.eq}) are defined in the
standard system; consequently, the equations must be applied iteratively
to measures from the {\sl WFPC2} frames. 

Based on both ground-based and {\sl HST} observations, emission lines
are not strong in regions B1 and B2, although some [O {\sc ii}]
$\lambda$3727 emission is present throughout and there are faint compact
H$\alpha$ sources in the western half of B2 (OM, de Grijs et al.\ 2000). 
We therefore do not include any correction for contamination by
emission-line flux of the optical standard magnitudes. 

The corresponding flight-to-{\it JH} transformation coefficients for the
calibration of our {\sl NICMOS} images, for $0.2 < (J-H) < 1.1$, were
adopted from Stephens et al.\ (2000):
\begin{eqnarray}
J =  -2.5 \times \log \dot{C}({\rm F110W})  -
(0.344 \pm 0.063) & &   \nonumber \\
\times (m[{\rm F110W}] - m[{\rm F160W}]) + (22.054 \pm 0.034) \, ,
\end{eqnarray}
and
\begin{eqnarray}
H = -2.5 \times \log \dot{C}({\rm F160W})  -
(0.305 \pm 0.065)  & &  \nonumber \\
\times (m[{\rm F110W}] - m[{\rm F160W}]) + (21.715 \pm 0.037) \, ,
\end{eqnarray}
where $m(i) = -2.5 \times \log \dot{C}(i)$.  [For their calibration,
Stephens et al.\ (2000) measured $\log \dot{C}(i)$ for calibrator stars
in an 0\asec5 aperture.] The colors of the cluster candidates in M82 B
fall within the range for which this calibration is applicable. 

We estimated foreground extinction by our Galaxy from Burstein \& Heiles
(1984), who give $A_B = 0.12$ mag for M82.  We determined the extinction
in the {\sl WFPC2} passbands assuming the Galactic extinction law of
Rieke \& Lebofsky (1985; $A_B / A_V = 1.324$), with the following
results: $A_{\rm F439W} / A_V = 1.367$, $A_{\rm F555W} / A_V = 1.081$,
$A_{\rm F814W} / A_V = 0.480$, $A_{\rm F110W} / A_V = 0.352$, and
$A_{\rm F160W} / A_V = 0.191$ mag. 

As a photometric consistency check, we compared our photometry of the
M82 B1 region with ground-based observations.  The total {\it V}-band
magnitude and $(B-V)$ color in a $7''$ diameter aperture, aligned with
the relatively sharp dust feature at the outer edge of the B1 region,
and not corrected for either foreground or internal extinction, are
$V_{\rm B1} = 13.3 \pm 0.2$; $(B-V)_{\rm B1} = 0.82$.  This corresponds
to a {\it V}-band surface brightness of $\mu_{V,\rm B1} = 17.3 \pm 0.2$
mag arcsec$^{-2}$.  Ground-based observations of the same area yield
$V_{\rm B1} = 13.05$ and $(B-V)_{\rm B1} = 0.84$ (OM).  Considering the
possible effects of centering differences and small atmospheric
zero-point offsets, the agreement is excellent.  In addition, comparison
of derived {\it V}-band magnitudes of isolated stars in our $\omega$ Cen
control field (\S \ref{control.sect}) with those obtained by D'Cruz et
al.\ (2000) from the same field showed agreement to within the
measurement errors. 

\subsection{Synthetic Control Fields}
\label{control.sect}

Small-scale variability in the bright M82 background is a serious
problem for our resolved-source photometry (see \S \ref{background.sect}
and Table \ref{background.tab}).  To assess its impact, we tested our
reduction and calibration procedures on synthetic data fields in two
different ways. 

{\bf (i)} First, we created artificial star fields by randomly and
uniformly adding 500 Gaussian star profiles to empty frames of the same
size as our science observations (with zero background).  We required
that the centers of the artificial sources should be at least 6 pixels
(0\asec273) from the edges of the frames.  The FWHMs of the synthetic
point sources were obtained from observational PSFs in the {\sl WFPC2}
PSF library, with FWHM$_{\rm F555W} = 1.44$ pixels or 0\asec066,
FWHM$_{\rm F814W} = 1.59$ pixels or 0\asec072, and they were scaled to
the desired magnitudes.  The histogram of input magnitudes was chosen to
approximate a power-law distribution, similar to those of the luminosity
functions in the M82 B regions, and extending to well below the 50\%
completeness limits (\S \ref{compl.sect}).  To simulate the
observational situation as closely as possible, we added Poisson noise
to the point sources.  We then added these control fields to the
observed M82 data frames to produce synthetic data frames to use as
tests of our photometric accuracy and completeness. 

{\bf (ii)} In \S \ref{starclussep:sect} we use image sizes to
discriminate stars from clusters.  So that we could test our technique
on stellar images which were as realistic as possible, we created
another set of synthetic images based on real {\sl HST} observations in
the F555W passband of an uncrowded star field in the globular cluster
$\omega$ Centauri obtained by D'Cruz et al.\ (2000).  We combined the PC
exposures of their ``position 2'' (J2000: RA = $13^{\rm h}26^{\rm
m}41.^{\rm s}48$, Dec = $-47^{\circ}31\arcmin 07$\asec38) into a single
median-filtered F555W image.  Sky background is near zero on this image. 
We then added this ``$\omega$ Cen field'' image, with different scale
factors, to our observed M82 data frames as a test of our techniques to
distinguish stars from clusters. 

In Fig.\ \ref{errors.fig} we show the error curves resulting from
measuring the brightnesses of the artificial sources in our type-(i)
control field in the same fashion as those of the real sources discussed
in \S \ref{sources.sect}.  As a function of input source magnitude we
have plotted the ($1 \sigma$) standard deviations of the distribution of
the differences between the input and retrieved magnitudes, for all
passbands and both regions.  The resulting error estimates (e.g.\
$\sigma_V \sim 0.1$ mag at $V \sim 21$) are significantly higher than
would be the case for isolated point sources on a clean sky background. 
The increase is due to the bright and variable background in M82. 

\section{General Morphology of Region B}
\label{morph.sect}

The mosaics for regions B1 and B2 resulting from our reduction are shown
in Figs.  \ref{B1.fig} and \ref{B2.fig}, respectively.  Each figure
includes the entire {\it V}-band {\sl WFPC2} mosaic for that region,
corresponding to the contours overlaid on the ground-based image (Fig.\
\ref{ground.fig}), surrounded by the full-resolution PC and {\sl NICMOS}
images of the region in the five filters used.  Figs. 
\ref{B1BvsNIR.fig} and \ref{B2BvsNIR.fig} are enlargements of the B1 and
B2 fields, respectively, showing the {\it B}-band PC image on the left
and the mean of the {\sl NICMOS} {\it J} and {\it H} mosaics on the
right. 

Both regions are riddled with dust lanes which become more prominent at
shorter wavelengths.  The densest lanes are found in the southwest
quadrant of region B2, nearest the center of M82.  Here, they are
detectable even on the longest-wavelength ({\it H}-band) image. 
Extinction effects completely transform the appearance of region B2
between the {\it H} and the {\it B} bands.  The most luminous objects in
B2 are obscured in the {\it B}-band by the lanes in the lower right
quadrant.  The {\it H}-band image of B1 is nearly free of distinct dust
lanes and shows a partially-resolved, bright concentration of light
running along the major axis of the galaxy.  This is undoubtedly the
underlying stellar disk of M82.  A dust lane bisects region B1.  The
photometry and spectra of region B discussed by OM and Marcum \&
O'Connell (1996) were based on observations of the easternmost part of
B1 (the upper left lobe in Fig.  \ref{B1BvsNIR.fig}). 

There are bright resolved or semi-resolved structures in all bands,
though fewer are present at short wavelengths.  The brighter of these
appear to be SSCs.  Only a few such objects are detectable even with
good seeing on ground-based images (e.g.\ object H in Fig.~1).  The
myriad of faint point sources detectable in the {\it I} band but more
easily visible on the {\it J} and {\it H} frames are luminous, cool
giants, resolved for the first time in the main body of the galaxy on
these exposures (\S \ref{stars.sect}).  The red giant branch tip was
first detected in the halo of M82 on {\sl HST} {\it V} and {\it I}
exposures at $I \sim 24.0$ by Sakai and Madore (1999). 

\section{Region B Background Surface Brightness}
\label{background.sect}

To determine the background surface brightness level in regions B1 and
B2, we sampled fluxes in 500 randomly placed 10-pixel radius apertures
in each field and filter.  Results are summarized in Table
\ref{background.tab}.  The modal background surface brightnesses and
dispersions are derived from Gaussian fits to the resulting flux
histograms.  The {\it B}-band value for region B1 is significantly
fainter than that quoted in \S \ref{calib.sect} because the $7''$
apertures used there were centered on the brightest subregion within the
B1 field. 

Both regions are anomalously bright compared to normal galaxies.  The
background surface brightnesses in Table \ref{background.tab} are $\sim
1 - 3$ mag arcsec$^{-2}$ brighter than the typical disk central surface
brightnesses of normal spiral galaxies, seen either face-on (e.g.\ de
Jong 1996) or edge-on (e.g.\ de Grijs 1998).  In fact, they are similar
to the bulge core brightnesses of normal galaxies.  As argued in \S
\ref{intro1.sect}, this is evidence for unusually intense star formation
in the past, extending up to a radius of 1 kpc, well beyond the confines
of the active present-day starburst. 

M82 would be regarded as a very unusual galaxy even in the absence of
its emission line plume and the radio/IR indicators of activity because
of the strange, high surface brightness optical structures found in its
central disk. 

\section{Cluster Identification and Photometry}

\subsection{Source Selection}
\label{sources.sect}

We based our initial selection of source candidates on a modified
version of the {\sc daofind} task in the {\sc daophot} software package
(Stetson 1987), running under {\sc idl}.\footnote{The Interactive Data
Language (IDL) is licensed by Research Systems Inc., of Boulder, CO.}
Experiments revealed a large disparity in candidate lists based on the
different wavelength bands, as might be expected from the appearance of
Figs.  \ref{B1.fig}--\ref{B2BvsNIR.fig}.  Candidates identified in the
near-IR images are often too faint for meaningful measurements in the
{\it B} or {\it V} bands and vice-versa. 

Therefore, we decided to combine and cross-correlate source lists
obtained in individual passbands and to supplement the automated
identifications with visual inspections.  We performed extensive checks
to find the best selection criteria and thus to minimize the effects of
dust features, residual cosmic rays, Poisson noise in regions of high
surface brightness, and the CCD edges.  We chose our detection
thresholds such that the numbers of initial candidates selected from
each passband were comparable.  Next, we cross-correlated the source
lists obtained in the individual passbands, allowing for only a 1-pixel
positional mismatch between the individual optical or near-IR bands; we
allowed for a 2-pixel mismatch when cross-correlating optical/near-IR
source positions, since the {\sl NICMOS} and {\sl WFPC2} optical paths
are slightly different.  This procedure led us to conclude that the
cross-correlation of the identifications in the F555W and F814W filters
({\it V} and {\it I\/}) contains the most representative fraction of the
population of genuine sources in M82 B. 

To reject artifacts remaining in the cross-correlated $V \otimes I$ list
and real sources which were badly situated for aperture photometry, we
visually examined all the candidates on enlargements of the F555W and
F814W images.  We discovered that many of the listed candidates were
artifacts and, moreover, that a number of plausible candidates had not
been included by the automated algorithms because of complicated
structures, extended profiles, or a highly non-uniform background. 
There were 268 and 147 sources in the B1 and B2 fields, respectively,
that were missed by the automated detection routine in F555W and F814W. 
We added these to the lists.  The complete source lists thus obtained
contain 735 and 640 candidates in M82 B1 and B2, respectively. 

Several systematic biases are inevitably introduced by the source
selection procedure.  Very blue or very red objects could be excluded
because of the primary $V \otimes I$ selection.  From the standpoint of
age-dating the star forming activity in region B, the blue omissions are
probably more important. 

Visual inspection of the {\it B}-band frames indicated that a number of
the obvious sources brighter than $B \simeq 24.0$ mag were missed.  We
added these sources to the lists, 40 and 25, respectively, in B1 and B2. 
In addition, most of the swarm of faint pointlike sources present in the
F110W and F160W frames are not detected in {\it V} either and therefore
are not included in our final source list.  Although most of these
fainter near-IR sources appear to be cool stars (we comment on them
separately in \S \ref{stars.sect}), we added the 58 and 94 brightest
near-IR sources that were missed by our automated detection routine to
the source lists for B1 and B2, respectively. 

Extinction is obviously a limiting factor in our ability to sample the
resolved sources in region B.  By basing our selection on {\it V} and
{\it I}, we are probably biased toward the surface regions of M82 B. 
However, we are confident that we have been able to detect nearly all of
the cluster candidates in the brightest 3--4 magnitudes of the
luminosity function in {\it any} of the observed passbands. 

The total number of visually verified sources contained in our source
lists is 833 and 759, respectively, for M82 B1 and B2. 

\subsection{Source Photometry}

The coordinates from the source lists obtained in the previous section
were used as the centers for {\sc daophot} aperture photometry in all
passbands. 

The correct choice of source and background aperture sizes is critical
for the quality of the resulting photometry.  Due to the complex
structure of M82 B, we concluded that we had to assign apertures for
source flux and background level determination individually to each
candidate by visual inspection.  The ``standard'' apertures for the
majority of the sources were set at a 5-pixel radius for the source
aperture and an annulus between 5 and 8 pixels for the background
determination, although in individual cases we had to deviate
significantly from these values. 

Our photometry includes most of the light of each cluster candidate.  At
the distance of M82, an aperture of radius 5 pixels (0\asec228)
corresponds to a projected linear diameter of $\sim 8$ pc, which is
larger than the sizes of the majority of SSCs that have been detected in
other galaxies (e.g.\ Whitmore et al.\ 1993, Hunter et al.\ 1994,
O'Connell et al.\ 1994, Barth et al.\ 1995, Whitmore \& Schweizer 1995,
Schweizer et al.\ 1996, Watson et al.\ 1996, Miller et al.\ 1997, De
Marchi et al.\ 1997).  Deconvolved {\sl HST WF/PC1}-images also indicate
that the young star clusters in the central regions of M82 typically
have half-peak intensity sizes of $\sim 3.5$ pc or 0\asec2 (O'Connell et
al.\ 1995). 

\subsection{Completeness}
\label{compl.sect}

We estimated the completeness of our source lists by using the synthetic
F555W and F814W images derived for artificial stars described in \S
\ref{control.sect}.  The artificial sources were given input magnitudes
between 20.0 and 26.0 mag and (F555W -- F814W) colors distributed around
zero ($V-I \approx 1.3$), which is approximately typical of the color
distribution of the cluster candidates (\S \ref{colors.sect}). 

We found that the effects of image crowding are small: only $\sim$
1--2\% of the simulated objects were not retrieved by the {\sc daofind}
routine due to crowding.  However, the effects of the bright and
irregular background and dust lanes are large.  Fig.\ \ref{compl.fig}
shows the fraction of simulated point sources that was recovered, as a
function of brightness in the {\sc stmag} system.  Incompleteness
becomes severe (i.e.\ the completeness drops below $\sim 50$\%) for
$m_{\rm ST}({\rm F555W}) > 23.1$ and 23.3 mag for M82 B1 and B2,
respectively, and for $m_{\rm ST}({\rm F814W}) > 23.5$ mag.  The
difference between F555W and Johnson {\it V} is F555W$-V \simeq 0.02$ to
0.05 for A0V to O5V spectral types, and between F814W and Cousins {\it
I} is F814W$-I \sim -1.1$ to $-1.2$ for similar spectral-type objects
(from the IRAF/{\sc synphot} package). 

Since the incompleteness threshold is mainly sensitive to the local
background noise, we added 200 synthetic sources to smaller areas
($250\times250$ pixels or 11\asec$375\times11$\asec375) in the B1 region
with relatively high and relatively low backgrounds in the F814W
passband.  The completeness curves resulting from this exercise are
shown in Fig.  \ref{compl.fig}c.  Our results are comparable to those
obtained from a similar exercise by Miller et al.\ (1997); they clearly
show the significant effects of the variable background. 

Foreground stars are not a source of confusion in the case of M82.  The
Milky Way models of Bahcall \& Soneira (1980) predict that only $1.5 \pm
0.2$ foreground stars, in the magnitude range of our cluster candidates,
would be in the PC field. 

\subsection{Separation of Clusters from Stars, Quality Assessment,
and Final Cluster Sample}
\label{starclussep:sect}

M82 is near enough, and the {\sl HST} images have high enough
resolution, that we can separate at least the larger clusters in M82
from stars on the basis of their sizes.  Separation on the basis of
brightness alone, which is the primary method available even with {\sl
HST} in the case of most putative SSC systems, is dangerous here because
of the possible presence of luminous supergiant stars. 

In general, the radial luminosity profile of a star cluster is
characterized by a core radius, $R_{\rm core}$, the radius at which the
surface brightness equals half the peak surface brightness, and an
effective radius, $R_{\rm eff}$, enclosing half of the total light.  In
practice, these radii are difficult to measure directly from {\sl HST}
observations because of resolution and background effects.  Therefore,
the method most often used to extract spatial information for marginally
resolved young star cluster candidates is to measure the magnitude
difference between 2 apertures, chosen to be characteristic of the
selected sources' spatial extents (e.g.\ Whitmore et al.\ 1993, Barth et
al.\ 1995, Whitmore \& Schweizer 1995, Holtzman et al.\ 1996, Schweizer
et al.\ 1996, Miller et al.\ 1997, Carlson et al.\ 1998).  These
magnitude differences can be calibrated by convolving the theoretical or
observational PSF with an assumed radial light distribution for the
clusters.  Unfortunately, the complex structure of M82 B precludes use
of this method.  We found from tests using our synthetic star fields (\S
\ref{control.sect}) that a variable background significantly distorts
the histogram of magnitude differences in the 2 aperture method,
although a smooth background does not.  This leads to overestimates of
the sizes of point sources. 

Instead of the aperture method, we have based our discriminant for
extended sources in the M82 B fields on the statistical differences
between the size characteristics of sources in these fields and those of
the stars in the $\omega$ Cen control field (\S \ref{control.sect})
added to the M82 B fields with scaling such that the output magnitudes
of the $\omega$ Cen stars were in the same range as those of the M82 B
sources.  We determined characteristic sizes in the F555W images of both
our M82 B candidate source list and the $\omega$ Cen stars using a 2-D
Gaussian fitting routine.  Although the true luminosity profiles of the
star clusters in M82 B may differ from Gaussians (\S \ref{sizes.sect}),
this method allows us to distinguish satisfactorily between compact and
extended sources.  Results of such fits to the candidate sources and to
the $\omega$ Cen stars injected into the M82 B images are shown in Fig. 
\ref{sigma.fig}.  The shaded histograms show the distribution of sizes
for the $\omega$ Cen control stars, while the open histograms show the
distributions for the candidate clusters in M82 B.  There is a clear
difference, demonstrating that many of the candidates are indeed
extended sources.  On statistical grounds, relatively little
contamination of our results by point sources will occur if we adopt for
our primary cluster sample objects with $\sigma_{\rm G} \ge 1.25$ pixels
(0\asec0569, or 1.0 pc), where $\sigma_{\rm G}$ is the best-fit Gaussian
sigma. 

We are mainly interested in obtaining good photometry for a
representative, if not complete, sample of M82 B clusters.  Therefore,
we visually inspected all candidates with $V \le 22.5$, i.e.\ slightly
brighter than the 50\% completeness limits, for contrast, definition,
aperture centering, and background sampling.  We rejected candidates
that were too diffuse or that might be the effects of background
fluctuations.  Many sources contained multiple components, and the
apertures were adjusted to include all of these.  We also visually
inspected the 2-D Gaussian fits to those sources that were assigned
$\sigma_{\rm G} \ge 2$ pixels in the initial automated fitting pass.  If
needed, apertures for our final photometric pass (and thus the source
magnitudes), center coordinates, selection criteria for inclusion into
the final source lists, and $\sigma_{\rm G}$'s were adjusted. 

Thus, our primary cluster candidate samples consist of well-defined
sources with $\sigma_{\rm G} \ge 1.25$ pixels, $V \le 22.5$ mag, and
relatively smooth backgrounds.  These samples contain 43 and 70 cluster
candidates in B1 and B2, respectively.  Their positions, sizes and
brightnesses are tabulated in, respectively, Tables \ref{srcB1.tab} and
\ref{srcB2.tab}.  The corresponding apparent {\it V}-band brightness
distribution for the cluster sample (without extinction corrections) is
shown in Fig.\ \ref{maghistHST.fig}.  Upper limits in any passband were
obtained by taking the brightest flux measurement in a 5-by-5 position
grid (within $25\times25$ pixels, or 1\asec14$\times$1\asec14). 

Fig.  \ref{VvsSigma.fig} shows our selection criteria in the $V$--
$\sigma_{\rm G}$ plane, where we have plotted all of the source
candidates regardless of quality, brightness, or size.  The horizontal
plume of sources at small sizes and faint magnitudes represents the
background of stars in the galaxy.  The brightest unresolved sources are
detected at $V \sim 21$--$22$ ($M_V \sim -6$).  This is consistent with
the expected brightness of luminosity class I supergiants in M82 before
extinction corrections (e.g.\ Humphreys \& McElroy 1984).  Few of the
sources in the primary cluster candidate lists are likely to be stars. 

\section{The Stellar Background in the Disk of M82}
\label{stars.sect}

As is evident from Fig.  \ref{VvsSigma.fig} or a close examination of
the original images, the galactic background of the M82 B regions is
predominantly composed of faint point sources, and these become
increasingly dominant at longer (near-IR) wavelengths.  Although Sakai
\& Madore (1999) detected individual giants in the halo of M82, this is
the first resolution of its disk into stars. 

In order to only select well-defined, real point sources from our
near-IR images, we required them to have matching detections in the {\it
I, J} and {\it H} bands.  Total magnitudes were determined from aperture
photometry, using aperture radii corresponding to 3 PC pixels;
background levels were measured in annuli with radii between 5 and 8
pixels.  Aperture corrections, based on synthetic {\sl HST} PSFs, were
applied. 

The near-IR brightness histograms of these faint point sources, which
are characterized by $\sigma_{\rm G} \approx 0.9 - 1.0$ pixels ($\sim
0$\asec040), peak around $J \approx 21$ and $H \approx 20$ mag.  Fig.\
\ref{NIRstars.fig} shows the corresponding color-absolute magnitude
diagrams for the $(I \otimes J \otimes H)$ cross-correlated point
sources with the highest-quality photometry (photometric uncertainty
$\delta(J-H) \le 0.15$ mag).  Sources in B1 and B2 are combined in the
plots; no extinction corrections are made. 

Overplotted in Fig.\ \ref{NIRstars.fig} are solar-metallicity stellar
isochrones from the Padova library corresponding to ages of 10, 30, and
100 Myr, with initial stellar masses ranging from 0.15 to $\sim 20, 9,$
and $5 M_\odot$, respectively (Girardi et al.\ 2000, Salasnich et al.\
2000).  In Figs.\ \ref{NIRstars.fig}c and d, we plot the two youngest
isochrones (10 and 30 Myr) on the same scale as in Figs.\
\ref{NIRstars.fig}a and b but use symbols whose sizes are proportional
to the predicted number of stars at each point, derived from the
cumulative mass functions tabulated with the isochrones. 

For comparison, we have added to the plots photometry for the late-type
K and M supergiant candidates in the Large Magellanic Cloud compiled
from earlier studies by Oestreicher, Schmidt-Kaler \& Wargau (1997)
(open circles).  These are clearly offset from the bulk of the M82 disk
stars.  The LMC supergiants are well modelled by massive stars with ages
of order 10 Myr.  The population in the M82 disk is evidently older. 
The isochrones indicate that the brightest M82 giants are core
helium-burning stars with ages $\sim 20-30$ Myr.  Significant star
formation has not occurred in region B in the last 10--15 Myr.  The age
estimates would not change significantly for metal abundances of
0.4--1.2 $Z_{\odot}$, the likely range for M82. 

Although the {\it J} and {\it H} bands are not very sensitive to dust
(cf.  the $A_V$ arrows in Fig.\ \ref{NIRstars.fig}), the few data points
scattered on the red side of the isochrones are most likely produced by
variable internal extinction within the disk of M82.  Our magnitude
cut-off prevents us from detecting the upper main sequence or giants
older than about 80 Myr. 

An age for the youngest bright giants in region B of $\gtrsim 20$ Myr is
consistent with the age estimates of the youngest clusters in our
cluster sample (see \S \ref{conspiracy.sect}), although the peak of
cluster formation was much earlier.  Only 4.3\% of the integrated {\it
J} band light in region B1 originates in the resolved population,
meaning that the great majority of cool stars were formed at earlier
times.  We return to the question of the star formation history in the
disk in \S \ref{summ.sect}. 

\section{Extinctions and Ages for Individual Clusters}
\label{populations.sect}

To examine the cluster luminosity function and the history of cluster
formation, we need reliable estimates for extinction within M82.  It is
obvious from the dust lanes and color contrasts within region B (see \S
\ref{morph.sect}) that we cannot simply adopt a mean value for the
internal extinction.  Instead, we need to estimate the extinction for
each cluster individually.  Fortunately, our multi-color imagery permits
a simultaneous estimate of cluster extinctions and ages through
comparisons to the colors of single-generation stellar populations
predicted by theoretical evolutionary spectral synthesis models. 

The set of such models most commonly used was developed by Bruzual \&
Charlot (1996, hereafter BC96, and references therein).  We find that
the more recently developed PEGASE models (Fioc \& Rocca-Volmerange
1997), which ought to be better suited for the study of young star
clusters since they include improved treatment of supergiant stars
(Gallagher \& Smith 1999), are roughly equivalent to BC96 for the colors
used in this paper.  Consequently, we have adopted the BC96 models.  We
use single-generation models with the Salpeter (1955) IMF and solar
metallicity ($Z_\odot \simeq 0.02$), unless otherwise indicated. 
Near-solar metallicity should be a reasonable match to the young objects
in M82 (e.g.\ Gallagher \& Smith 1999).  Fritze-v.\ Alvensleben \&
Gerhard (1994) also find from chemical evolution models that young
clusters should have $Z \sim 0.3 - 1.0 Z_\odot$. 

\subsection{Age and Extinction Separation in the Two-Color Diagram}
\label{conspiracy.sect}

In Fig.\ \ref{colcol.fig} we present a two-color diagram for the cluster
samples in M82 B1 and B2.  Overplotted with the thick lines are the
colors for solar-metallicity BC96 single-generation models from 10 Myr
to 20 Gyr.  Reddening trajectories for selected models (ages of 10 Myr,
100 Myr, 1 Gyr, and 10 Gyr) are shown with thin lines for a foreground
screen extinction geometry; these are crossed by thin lines
corresponding to $A_V =$ 1, 2, and 3 magnitudes. 

The clusters in the diagram are distinguished as a function of source
brightness: the filled circles represent sources with $V \le 21.0$,
while the open circles are the remaining sources brighter than $V =
22.5$.  This includes objects for which we only have upper limits in one
passband.  For these sources, the length of the arrow indicates the
measurement uncertainty resulting from the $5\times5$ position grid
sampling (\S \ref{sources.sect}). 

In Fig.\ \ref{colcol.fig} the observations fall, as they should, in the
``allowed'' part of the plot, near or above the locus of unextincted,
single-generation populations.  The few sources below the envelope are
consistent with the photometric errors.  The larger scatter and slightly
redder colors in region B2 imply that the sources in B2 are, on average,
more affected by extinction than those in B1. 

The model grid shows that a {\it BVI} diagram is well suited to
disentangle the effects of age and extinction.  We explored the use of a
similar {\it VIH} two-color diagram.  Unfortunately, the aging and
extinction trajectories are nearly parallel here, and it is much less
useful for age/extinction separation.  The BC96 models, the PEGASE
models, and the Starburst99 models (Leitherer \& Heckman 1995, Leitherer
et al.\ 1999) all give similar results in the {\it VIH} plane. 
 
We can use the location of each cluster in Fig.\ \ref{colcol.fig} to
estimate its age and extinction, assuming solar metallicity.  The
effects of varying the metallicity from $0.2 Z_\odot$ to $2.5 Z_\odot$
for the BC96 models are $\lesssim 0.1$ mag in each of our colors, so
differences in metallicity will not strongly affect the derived age
distribution.  

Our best age and extinction estimates for each cluster, as well as the
extinction-corrected colors and absolute {\it V}-band magnitudes, are
listed in Tables \ref{derivedB1.tab} and \ref{derivedB2.tab} for B1 and
B2, respectively.  Typical estimated errors in the ages are $\pm 40$\%. 
Our treatment ignores the effects of any extinction internal to the
clusters themselves. 

\subsection{Color Comparison to Other Cluster Systems}
\label{colors.sect}

In Fig.\ \ref{cluscols.fig} we show the $(B-V)_0$ and $(V-I)_0$ color
histograms of our cluster candidates, corrected for internal extinction
on a cluster-by-cluster basis (\S \ref{conspiracy.sect}).  We also show
the corresponding color distributions of comparison samples of SSC
systems based on {\sl HST} photometry.  These are corrected for
foreground (Galactic) extinction but not for extinction within the
parent galaxy.  In Table \ref{colpeaks.tab} we have collected the
characteristics of the color distributions of these cluster samples,
including their estimated ages. 

It is immediately clear that the M82 B color distributions are redder
than those of the comparison systems.  Since the latter reside in
regions with high and non-uniform dust extinction in their parent
galaxies (e.g.\ Whitmore et al.\ 1993, Whitmore \& Schweizer 1995,
Miller et al.\ 1997, Carlson et al.\ 1998), their intrinsic colors are
likely even bluer.  The M82 A cluster sample is corrected only for
foreground Galactic reddening, and the extremely red colors are due to
the very high extinction in the active starburst region (O'Connell et
al.\ 1995). 

The M82 B clusters are intrinsically both redder and (see \S
\ref{clf.sect}) less luminous (because of fading with age) than those in
the ``classic'' SSC systems.  We were able to survey this system well
only because of its proximity.  This emphasizes the strong selection
effects which operate, even with {\sl HST}, for identification of
clusters in more distant galaxies. 

\subsection{The Cluster Formation History of M82 B}
\label{sfh.sect}

In Fig.\ \ref{ages.fig} we show the age distributions for the clusters
in regions B1 and B2.  Both contain clusters with a wide range of ages,
from $\sim 30$ Myr to over 10 Gyr.  The distributions are statistically
identical.  In each case, about 22\% of the clusters are older than 2
Gyr, with a flat distribution to over 10 Gyr.  There is a strong peak of
cluster formation at $\sim 600$ Myr ago but very few clusters are
younger than 300 Myr.  The full-width of the peak is $\sim$ 500 Myr, but
this is undoubtedly broadened by the various uncertainties entering the
age-dating process.  The selection bias is such that the truncation of
cluster formation for $t < 200$ Myr is better established than is the
constant formation rate at $t > 2$ Gyr. 

The photometry suggests steady, continuing cluster formation at a very
modest rate at early times ($> 2$ Gyr ago) followed by a concentrated
formation episode lasting from 400--1000 Myr ago and a subsequent
suppression of cluster formation.  Region B has evidently not been
affected by the more recent ($< 30$ Myr) starburst episode now
continuing in the central regions. 

It has long been supposed that tidal interactions among the nearby M81
group members are responsible for the unusual gas streamers in the group
and the starburst in M82 (Gottesman \& Weliachew 1977, Cottrell 1977,
van der Hulst 1979, Yun 1992, Yun et al.\ 1994).  The observed
distribution of intragroup gas is consistent with a 3-body model in
which there was a perigalactic passage between M82 and M81 (at a
distance of 21 kpc) 500 Myr ago (Brouillet et al.\ 1991).  This
independent dynamical estimate of the last M81/M82 passage is remarkably
close to the peak of the cluster formation burst seen in our data.  This
suggests that cluster formation was induced in the disk of M82 by the
last encounter with M81.  The current starburst in the center of the
galaxy is probably related to late infall of tidally disrupted debris
from M82 itself (OM, Yun, Ho, \& Lo 1993). 

\section{Estimated Cluster Masses}
\label{masses.sect}

We present the distribution of extinction-corrected absolute magnitudes
of the M82 B cluster sample in panels {\it (a)} and {\it (b)} of Fig.\
\ref{absv.fig}.  In Fig.\ \ref{lumdist.fig}a we display the
corresponding {\it V}-band intrinsic luminosity function for regions B1
and B2 combined.  Using the age estimates derived in \S
\ref{populations.sect}, we can now apply the age-dependent mass-to-light
ratio predicted for a single burst stellar population by BC96 to derive
estimated masses for our cluster sample.  Our estimates assume a
Salpeter (1955) initial mass function.  The results are shown in Fig.\
\ref{lumdist.fig}b.  The masses of the young clusters in M82 B with $V
\le 22.5$ mag are mostly in the range $10^4 - 10^6 M_\odot$, with a
median of $10^5 M_\odot$. 

The high end of the M82 cluster mass function overlaps with those
estimated by similar techniques for young SSCs in other galaxies (e.g.\
Richer et al.\ 1993, Holtzman et al.\ 1996, Tacconi-Garman, Sternberg,
\& Eckart 1996, Watson et al.\ 1996, Carlson et al.\ 1998).  Independent
dynamical mass estimates are available only for a few of the most
luminous SSCs, one of them M82 F, and are approximately $10^6 M_\odot$
(Ho \& Filippenko 1996a,b; Smith \& Gallagher 2000).  Because of the
proximity of M82, we have been able to probe the young cluster
population in M82 B to fainter absolute magnitudes, and thus lower
masses, than has been possible before in other galaxies.  Other SSC
cluster samples are biased toward high masses by selection effects. 

The M82 cluster masses are comparable to the masses of Galactic globular
clusters (e.g.\ Mandushev, Spassova, \& Staneva 1991, Pryor \& Meylan
1993), which are typically in the range $10^4$ -- $3 \times 10^6
M_\odot$.  If they survive to ages of $\gtrsim 10$ Gyr, the M82 clusters
will have properties similar to those of disk population Galactic
globulars. 

\section{The Cluster Luminosity Function}
\label{clf.sect}

The cluster luminosity function (CLF) is one of the most important
diagnostics in the study of globular and SSC populations.  For the old
globular cluster systems in, e.g.\ the Galaxy, M31, M87, and old
elliptical galaxies, the CLF shape is well-established: it is roughly
Gaussian\footnote{Note that fitting the CLF shape with a Gaussian
distribution function has been chosen only for mathematical convenience;
any physical rationale for a Gaussian distribution is currently lacking
(see also Harris, Harris, \& McLaughlin 1998).}, with the peak or
turnover magnitude at $M_V^0 \simeq -7.4$ and a Gaussian FWHM of $\sim
3$ mag (Harris 1991, Whitmore et al.\ 1995, Harris et al.\ 1998). 
Harris (1996) has shown that this turnover magnitude depends only weakly
on galaxy luminosity and type. 

The well-studied young star cluster population in the LMC, on the other
hand, displays a power-law CLF of the form $\phi_{\rm young}(L) {\rm d}
L \propto L^{\alpha} {\rm d} L$.  where $\phi_{\rm young}(L) {\rm d} L$
is the number of young star clusters with luminosities between {\it L}
and $L + {\rm d} L$, with $-2 \lesssim \alpha \lesssim -1.5$ (Elson \&
Fall 1985, $\alpha = -1.5 \pm 0.2$; Elmegreen \& Efremov 1997, $\alpha
\ge -2$). 

{\sl HST} observations have provided CLFs for young, compact cluster
systems in more distant galaxies.  Examples, including our M82
observations, are shown in Fig.\ \ref{absv.fig}, where we include the
CLF of the globular cluster system in our Galaxy for reference.  Where
available, we have indicated the completeness limits by dashed lines. 
Although incompleteness effects often preclude detection of a turnover
in the CLF at the expected magnitude (Whitmore \& Schweizer 1995, NGC
4038/39; Schweizer et al.  1996, NGC 3921; Miller et al.\ 1997, NGC
7252), in galaxies for which deep {\sl HST} observations are available
(e.g.\ NGC 1275, Carlson et al.\ 1998; NGC 7252, Whitmore et al.  1993)
there is no strong evidence for an intrinsic turnover.  The CLF shapes
are consistent with power laws down to the completeness threshold (but
see Miller et al.\ 1997). 

The striking differences between the power-law SSC distributions and the
Gaussian distribution of the old Galactic globulars has recently
attracted renewed theoretical attention.  Globular cluster formation
models suggest that the distribution of the initial cluster masses is
closely approximated by a power law of the form d$N(M) {\rm d}M \propto
M^{\alpha} {\rm d}M$, where $-2.0 \lesssim \alpha \lesssim -1.5$ (e.g.\
Harris \& Pudritz 1994, McLaughlin \& Pudritz 1996, Elmegreen \& Efremov
1997).  In fact, Elmegreen \& Efremov (1997) and Harris \& Pudritz
(1994) argue that the {\em initial} mass distribution functions for,
among others, young and old star clusters are universal, and independent
of environment.  Ostriker \& Gnedin (1997) agreed and argued that the
differences in shape between the inner and the outer globular cluster
luminosity functions in our Galaxy, M31 and M87 are solely due to
dynamical cluster evolution and not to intrinsically different initial
mass distributions.  McLaughlin, Harris, \& Hanes (1994) also concluded
that any differences in the M87 luminosity distribution as a function of
radius could be accounted for by dynamical evolution of the globular
cluster populations. 

Which processes will affect the CLFs such that they transform from a
power-law shape to a Gaussian distribution? It is generally assumed that
the processes responsible for the depletion of, preferentially,
low-luminosity, low-mass star clusters over time scales of a Hubble time
are tidal interactions with the background gravitational field of the
parent galaxy and evaporation of stars through two-body relaxation
within clusters (e.g.\ Fall \& Rees 1977, 1985, Elmegreen \& Efremov
1997, Murali \& Weinberg 1997a,b,c, Ostriker \& Gnedin 1997, Harris et
al.  1998, and references therein).  From the models of Gnedin \&
Ostriker (1997) and Elmegreen \& Efremov (1997) it follows that {\it
any} initial mass (or luminosity) distribution will shortly be
transformed into peaked distributions. 

We now consider the M82 B CLF in this context.  The combined CLF for M82
B1 and B2 is shown in Fig.\ \ref{lumdist.fig}a in units of solar
luminosity.  Although the M82 B CLF exhibits a turnover at $L_V^0 \simeq
4.92 L_{V,\odot}$, or $M_V^0 \simeq -7.5$, incompleteness likely affects
this result. 

For the proper interpretation of the M82 B CLF, we need to correct the
cluster luminosities for the large range in ages found in \S
\ref{populations.sect}.  Using the BC96 models, we have corrected the
present-day luminosities of our clusters to those at a fiducial age of
50 Myr, comparable to the ages of most young SSC populations in other
galaxies (cf.\ Table \ref{colpeaks.tab}).  The results are shown in
Fig.\ \ref{lumdist.fig}c. 

The constant-age CLF for the bright clusters in M82 B ($M_V^0[50]
\lesssim -9.75$) follows a power law with a slope of $\sim -1.2$.  We
have plotted the power law shapes for slopes of $\alpha = -1.2, -1.5$,
and $-2$ for reference in Fig.\ \ref{lumdist.fig}c.  The formal
uncertainty in our best-fitting power law solution ($\Delta \alpha
\simeq 0.3$) is such that an $\alpha = -1.5$ shape cannot be ruled out
at the 1$\sigma$ level, while an $\alpha = -2$ power law is highly
unlikely.  The bright cluster, constant-age CLF in M82 is therefore
consistent with the power-law shapes of other young star CLFs. 

The overall constant-age CLF for M82 B is, however, broader and flatter
than typical of the younger cluster systems seen in Fig.\
\ref{absv.fig}.  It more closely resembles what is expected for a
cluster system which has begun to undergo significant dynamical
evolution.  It would not be surprising if the depletion of low-mass
clusters in M82 was well underway.  Region B has a high density and has
been subject to the dynamical disturbances associated with the tidal
interaction with M81.  Fig.\ \ref{absv.fig} is, however, subject to
strong selection effects whose modeling is beyond the scope of this
paper. 

\section{Cluster Structures}
\label{sizes.sect}

Because of M82's proximity, we can study the structure of its compact
clusters better than is possible for other SSC systems.  Estimates of
star cluster sizes are heavily dependent on the radial brightness
distribution assumed (e.g.\ Holtzman et al.\ 1996).  Several functions
have been used to obtain size estimates, the simplest of them being a
two-dimensional Gaussian, which is characterized by $R_{\rm core} =
R_{\rm eff} =$ FWHM.  Even though mathematically convenient, a Gaussian
profile may not be the best representation of the radial luminosity
profiles of star clusters, since they may have more extended wings than
Gaussian distributions, thus causing an underestimate the true
half-light radii (see also Holtzman et al.\ 1996). 

Our cluster sample was defined (\S \ref{starclussep:sect}) to have
$\sigma_{\rm G} > 1.25$ pixels, measured in the {\it V} band.  Most of
the sample falls in the range $1.25 \le \sigma_{\rm G} \lesssim 5.5$
pixels (Fig.\ \ref{sigma.fig}), or $2.34 \le R_{\rm eff} \lesssim 10$
pc, with a wing extending to larger sizes.  Keeping in mind that, due to
our size selection limit, we are only probing the more extended star
clusters, these sizes are consistent with those obtained from {\sl HST}
imaging for other (more distant) young star cluster systems (e.g.\
O'Connell et al.\ 1994, Barth et al.\ 1995, Holtzman et al.\ 1996,
Schweizer et al.\ 1996, Miller et al.\ 1997, Whitmore et al.\ 1997) and
with the Galactic globular clusters (van den Bergh, Morbey, \& Pazder
1991, Djorgovski 1993, van den Bergh 1995).  Given their sizes and ages,
which amount to many dynamical crossing times (of typically a few Myr;
cf.  Schweizer et al.  1996) for their estimated masses, the M82
clusters are predominantly gravitationally bound. 

Alternatively, one can fit more complex models that are better
representations of the true stellar light distribution in compact
clusters. The most general of these is the mathematically convenient
function of surface brightness $\mu(r)$ as a function of radius {\it r},
proposed by Elson, Fall, \& Freeman (1987):
\begin{equation} 
\label{elson.eq}
\mu(r) = \mu_0 \Biggl( 1 + \Bigl( {r \over R_{\rm core}} \Bigr)^2
\Biggr)^{-\gamma/2} .
\end{equation} Equation (\ref{elson.eq}) reduces to a modified Hubble
law for $\gamma = 2$, which is a good approximation to the canonical
King model for globular clusters (King 1966). Elson et al.\ (1987)
found, for 10 rich star clusters in the LMC, that $2.2 \lesssim \gamma
\lesssim 3.2$, with a median $\gamma = 2.6$. Most of these clusters do
not seem to be tidally truncated, although cluster radial profiles are
often modeled as truncated modified Hubble laws, with the truncation
occurring at $\sim$ 10--1000 $R_{\rm core}$ (e.g.\ Holtzman et al.\
1996, Watson et al.\ 1996).

Due to low signal-to-noise ratios, complicated galactic backgrounds, and
the distance to most SSC systems, the fitting to determine the core
radii of SSCs has proven difficult.  Previous studies have only been
able to place upper limits on the core radii (Holtzman et al.\ 1996 [NGC
3597: $R_{\rm core} \lesssim 2$ pc], Carlson et al.\ 1998 [NGC 1275:
$R_{\rm core} \lesssim 0.75$ pc]). 

We fit modified Hubble profiles ($\gamma =2$) to the radial luminosity
distributions within the inner 3 pixels (2.4 pc) of the M82 B cluster
sample with $\sigma_{\rm G} > 1.25$ pixels, using a customized version
of the {\sc curvefit} subroutine running under {\sc idl}.  It employs a
gradient-expansion algorithm to compute a non-linear least squares fit
to a modified Hubble profile.  We find core radii of $3 \lesssim R_{\rm
core} \lesssim 10$ pixels, or $2.4 \lesssim R_{\rm core} \lesssim 7.9$
pc, with a smaller, secondary peak at $\sim 1.2$ pc.  These core radii
are similar to those for the Galactic globular clusters (Djorgovski
1993): $\langle R_{\rm core,G} \rangle \sim 1$ pc, with a total range
from 0.03 to 23.4 pc. 

In Fig.\ \ref{bright.fig} we present the {\it V}-band light profiles of
the five brightest star cluster candidates in each field.  Their
photometric characteristics and the size estimates obtained from fits to
the innermost 8 pixels are summarized in Table \ref{bright.tab}.  These
objects are bright and relatively isolated, thus enabling us to follow
the light profiles further out.  The modified Hubble profile fits are
generally better than the Gaussian fits.  The second and fourth columns
of Fig.\ \ref{bright.fig} show the flux ratio of the data points to both
the Gaussian and the modified Hubble profile approximations.  For
reasons of clarity we did not include the equivalent ratio for the
$\gamma = 2.6$ profile in the figure.  However, the residuals for the
$\gamma = 2.6$ profiles are similar to or slightly better than the
$\gamma = 2.0$ fits. 

To assess the robustness of our fitting procedures, we obtained
independent model fitting results based on the two-dimensional profile
fitting routines of Matthews et al.\ (1999), which take into account the
detailed {\sl HST} PSFs at the positions of our M82 B clusters.  L.D. 
Matthews kindly provided us with detailed profile fits for the brightest
clusters in each of our fields, and for each of the functionalities
assumed in this paper.  The difference between her and our results is
$\le 10-15$\%, and does not appear to be systematic. 

Some of the star clusters in Fig.\ \ref{bright.fig} show evidence for
asymmetrical structures, elongations or subclustering, which may mean
that they have not yet reached dynamical equilibrium.  We may be
witnessing ongoing merging processes, although we will need
spectroscopic follow-up observations to unambiguously determine this. 
Similar subclustering has been observed in the SSCs in NGC 1569
(O'Connell et al.\ 1994, De Marchi et al.\ 1997) and in some of the rich
LMC star clusters (e.g.\ Fischer, Welch, \& Mateo 1993). 

\section{Summary and Discussion}
\label{summ.sect}

We have presented high-resolution {\sl HST WFPC2} and {\it NICMOS}
imaging of two adjacent fields in the disk 1 kpc NE of the center of
M82, the prototype local starburst galaxy.  This region, M82 ``B,'' has
been suspected to be a site where an intense episode of star formation
occurred over 100 Myr ago.  Our new observations confirm that M82 B is a
``fossil'' of an ancient starburst with an amplitude entirely comparable
to that now transpiring in the core of M82.  It is an analog of the
anomalous E$+$A systems found in distant galaxy clusters.  M82 thus
presents us with the unique opportunity to observe both active and
evolved starburst environments at close range. 

The main results of our multi-passband study are summarized below:

\begin{itemize}

\item The surface brightness of M82 B is well above normal for galactic
disks and comparable to the core surface brightnesses in spiral
galaxies.  Its intrinsic surface brightness at an age of 10 Myr was
comparable to that found in the present-day nuclear starburst,
indicating an event of comparable amplitude. 

\item Because of the bright unresolved background light and the large
and highly variable extinction within M82, we have been forced to use
specialized methods to identify and photometer sources.  We tested
selection and photometry methods using artificial sources or real star
field images from {\sl HST} injected into our observed images.  The
proximity of M82, however, is a key advantage relative to studies of
more distant galaxies systems because we can use source sizes as a
criterion to distinguish stars from star clusters. 

\item We find a large, evolved system of SSCs in M82 B.  Our cluster
candidate samples consist of well-defined sources with Gaussian
$\sigma_{\rm G} \ge 1.25$ pixels (0\asec0569, or 1.0 pc), $V \le 22.5$
mag, and relatively smooth backgrounds.  These samples contain 43 and 70
cluster candidates in regions B1 and B2 (nearest the M82 nucleus),
respectively. 

\item We use a two-color {\it BVI} diagram and evolutionary spectral
synthesis models to separately estimate the extinction and age of each
cluster.  The clusters range in absolute magnitude from $M_V^0 = -6$ to
--10, with a peak at --7.5.  The derived age distribution suggests
steady, continuing cluster formation at a modest rate at early times ($>
2$ Gyr ago), followed by a concentrated formation episode $\sim 600$ Myr
ago and more recent suppression of cluster formation.  The peak episode
coincides with independent dynamical estimates for the last tidal
encounter with M81, which presumably induced the starburst. 

\item Our {\it J} and {\it H} band observations resolve the stellar
population in M82's disk for the first time.  Comparing our near-IR
color-magnitude diagram to recent Padova isochrones, the detected stars
appear to be cool, helium-burning bright giants.  Star formation
evidently continued in M82 B until about 20 -- 30 Myr ago.  However,
significant star formation has not occurred in M82 B in the last 10--15
Myr, during which the nuclear starburst has been highly active. 

\item After correcting the cluster luminosity function to a fiducial age
of 50 Myr, we find that the bright end is characterized by a power-law
slope with $\alpha = -1.2 \pm 0.3$, similar to that of other young
cluster systems in interacting galaxies.  There is tentative evidence
for broadening of the luminosity function due to dynamical destruction
of lower mass clusters. 

\item The radial luminosity profiles of the brightest clusters are more
closely approximated by modified Hubble or ``Elson'' functions with
$\gamma = $ 2.0 or 2.6 than by Gaussians.  Some of the clusters show
evidence for asymmetries or subclustering, possibly an indication of
mergers.  Cluster sizes ($2.34 \le R_{\rm eff} \lesssim 10$ pc, or $2.4
\lesssim R_{\rm core} \lesssim 7.9$ pc) and estimated masses (a median
of $10^5 M_\odot$) are consistent with values found for young super star
cluster populations in M82's core and other galaxies and with the
progenitors of globular clusters. 

\end{itemize}

When we combine the age estimates for the evolved cluster sample in M82
B, for the integrated spectrum of M82 B (Marcum \& O'Connell 1996), and
for the resolved cool giants in the disk of region B, the following
picture for the star formation history in its disk emerges. 

The last tidal encounter between M82 and M81 about 500 Myr ago had a
major impact on what was probably an otherwise normal, quiescent, disk
galaxy.  It caused a concentrated burst of star formation activity, as
evidenced by the peak in the age distribution of the cluster sample in
M82 B.  Comparison of the cluster ages with the integrated light dating
($\sim 100$--200 Myr) suggests that field star formation may have
continued at a high rate after cluster formation had begun to decline,
but the uncertainties in the methods are too large to be certain.  The
enhanced cluster formation decreased rapidly within a few hundred Myr of
its peak.  However, field star formation continued, probably at a much
lower rate, in M82 B until $\sim 20$ Myr ago.  It has evidently been
suppressed during the last $\sim 10 - 15$ Myr, during which the
starburst in the core of M82 has been most active.  Evidence for
supernova remnants in the parts of region B2 nearest the starburst core
(de Grijs et al.\ 2000) but not in region B1 indicates that disk star
formation during the last 50 Myr was more active nearer the nucleus. 
The current starburst is probably related to late infall of tidally
disrupted debris from M82 itself. 

The evidence for decoupling between cluster and field star formation is
consistent with the view that SSC formation requires special conditions,
e.g.  large scale gas flows, in addition to the presence of dense gas
(cf.  Ashman \& Zepf 1992, Elmegreen \& Efremov 1997). 

A strong tidal interaction could easily produce an off-nuclear starburst
at a site like M82 B.  However, the M82 B burst could also have been
part of a larger scale event encompassing the center of the galaxy as
well.  Given the high extinction and the dominance by much younger
concentrations of stars, it would not be easy to identify older clusters
in the starburst core if they exist. 

All of our observational evidence points to a scenario in which the
intermediate-age star cluster population in M82 B is entirely comparable
to SSC populations at a younger age, as seen in M82 A and other
galaxies.  They will likely evolve into Galactic globular cluster
analogs over a Hubble time. 

\paragraph{Acknowledgements} - We thank Noella D'Cruz for helping us to
check our optical photometry, Andrew Stephens and Jay Frogel for making
their {\sl NICMOS} calibration coefficients available prior to
publication, Allan Sandage for loan of the plate shown in Fig.  1, Lynn
Matthews for providing an independent check on our source size
estimates, M\'arcio Catelan for useful discussions about the
interpretation of the stellar disk light in M82, and Mark Whittle for
many insightful and stimulating discussions.  This research was
supported by NASA grants NAG 5-3428 and NAG 5-6403, and has made use of
NASA's Astrophysics Data System Abstract Service and of {\sl HST}
archival data at the STScI.

\newpage
\figcaption[RdeGrijs.fig1.ps]{\label{ground.fig}M82 {\it B}-band image,
extracted from a Palomar 5m plate taken by Sandage (exposure time 20
minutes, seeing $\lesssim 1''$; cf. O'Connell et al. 1995); the {\sl
WFPC2} fields covered by the observations presented in this paper are   
indicated. The locations of regions M82 A, C and H are also indicated.}

\figcaption[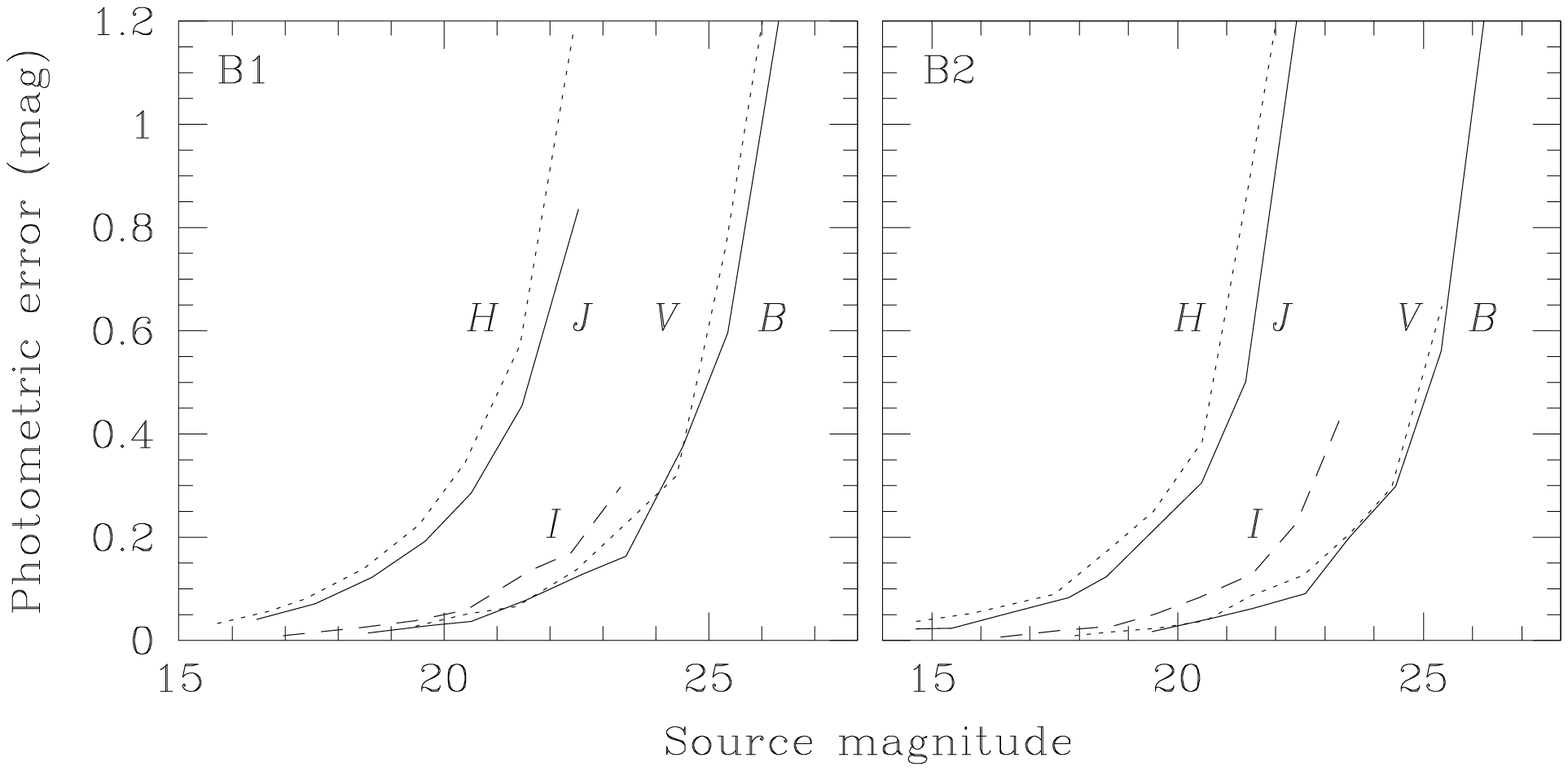]{\label{errors.fig}Photometric error curves
for the M82 B fields, as a function of input (synthetic) source
magnitude.}

\figcaption[RdeGrijs.fig3.ps]{\label{B1.fig}{\it V}-band {\sl WFPC2}
mosaic of M82 B1, and PC (or equivalent) fields for all passbands; the
{\sl NICMOS} fields were adjusted to match the PC field and pixel size.}

\figcaption[RdeGrijs.fig4.ps]{\label{B2.fig}Observations of M82 B2,
similar to those of B1 in Fig.  \ref{B1.fig}.}

\figcaption[RdeGrijs.fig5.ps]{\label{B1BvsNIR.fig}Comparison of the
detailed structures in M82 B1 between the optical {\it B} band and the
mean of the near-IR {\it J} and {\it H} filters.}

\figcaption[RdeGrijs.fig6.ps]{\label{B2BvsNIR.fig}Same as Fig. 
\ref{B1BvsNIR.fig}, but for region B2.}

\figcaption[RdeGrijs.fig7.ps]{\label{compl.fig}Results of completeness
tests obtained from artificial star counts as described in the text.}

\figcaption[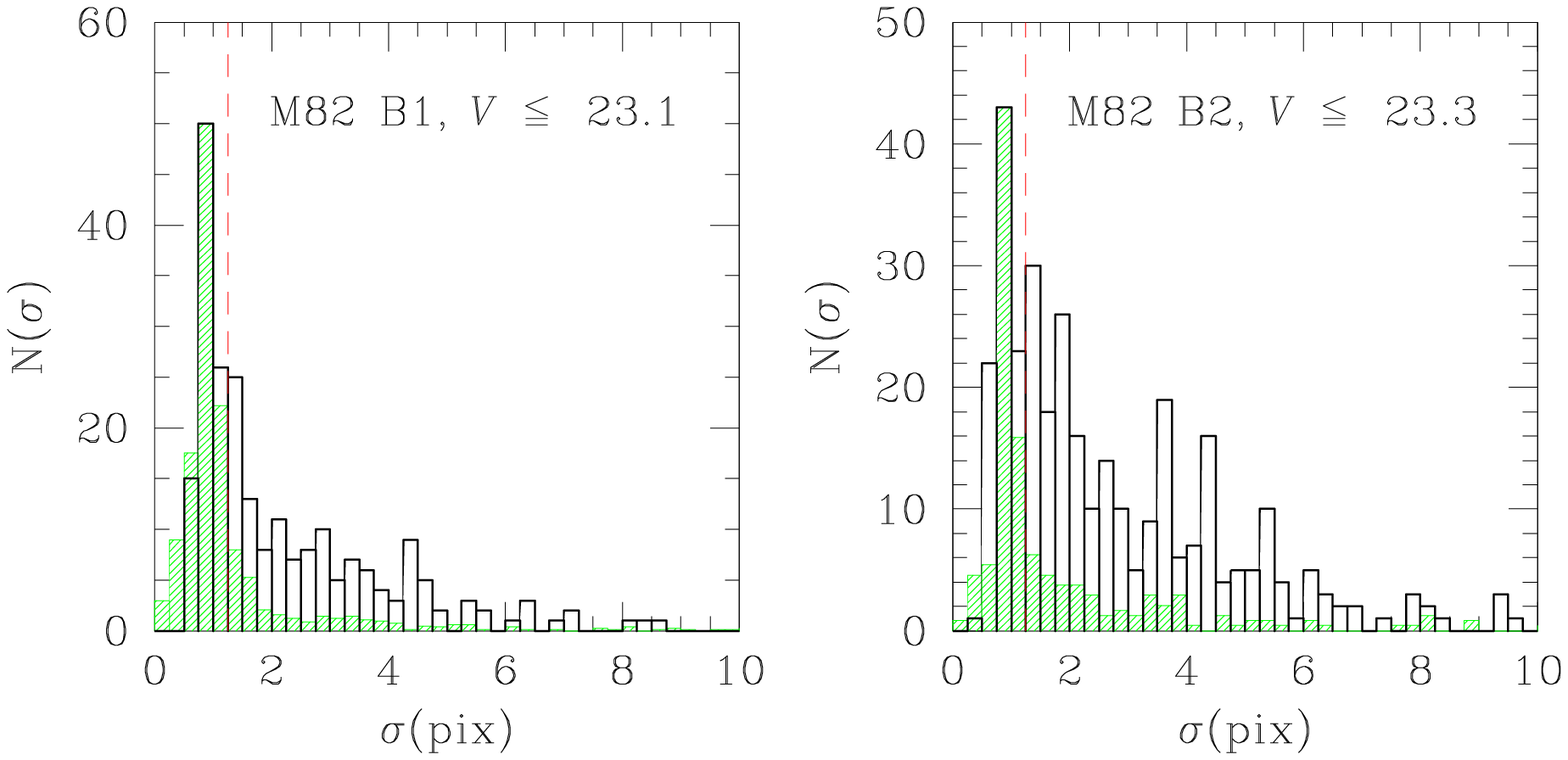]{\label{sigma.fig}Distribution of
$\sigma_{\rm Gaussian}$ for B1 and B2 (open histograms).  All sources
down to and including the 50\% completeness limits have been included. 
The shaded histograms are the corresponding size distributions for the
$\omega$ Cen control field, scaled to the peak of the M82 B histograms
and added to the observed fields.  We have indicated the size limit
adopted to distinguish between stars and star clusters by the dashed
line at $\sigma_{\rm G} = 1.25$ pixels.}

\figcaption[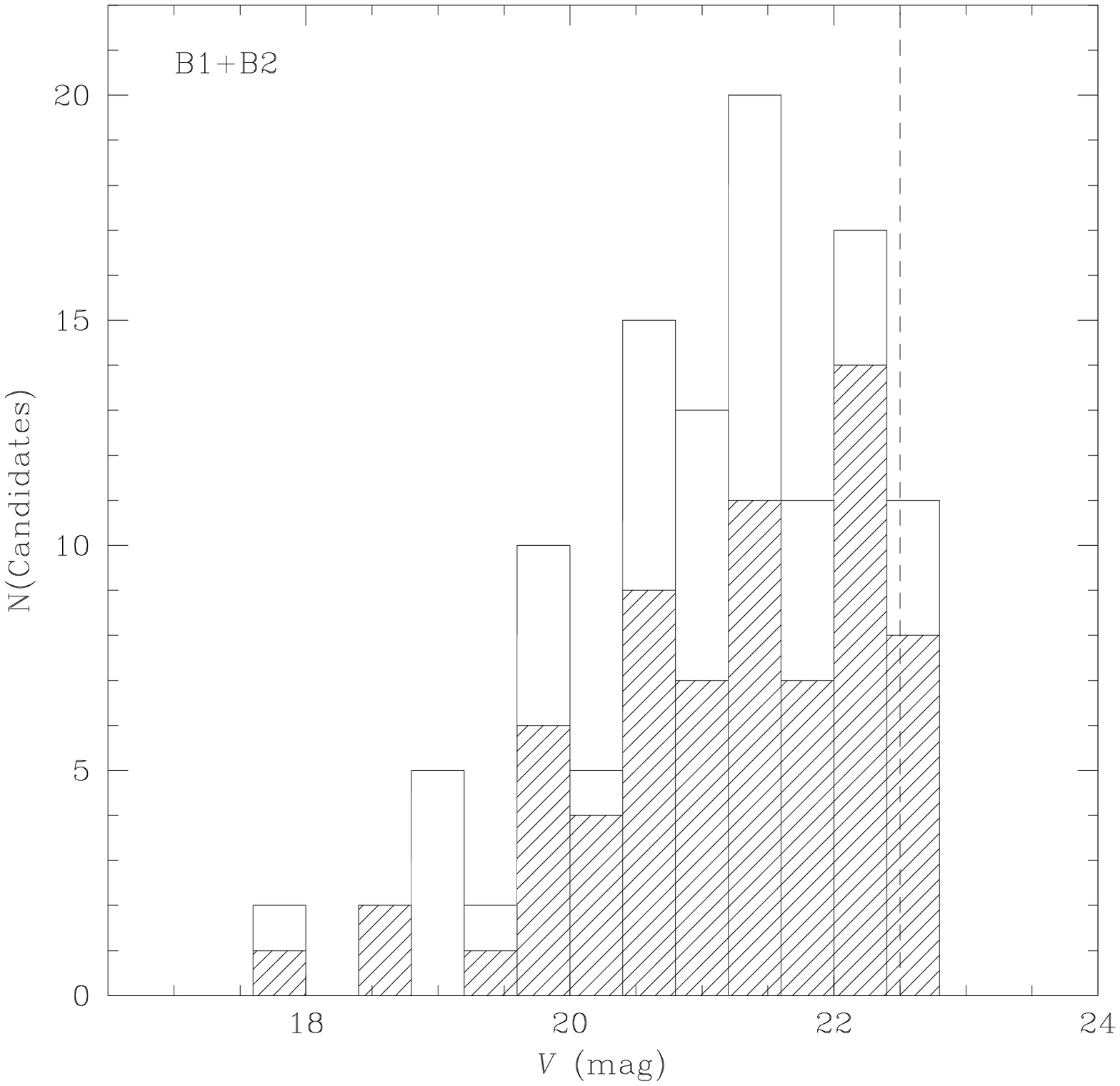]{\label{maghistHST.fig}Cluster luminosity
functions for our sample of candidate SSCs.  The open histogram
represents the CLF for M82 B1; the shaded histogram shows the B2 CLF. 
The dashed line indicates our selection limit at $V = 22.5$ mag.}

\figcaption[RdeGrijs.fig10.ps]{\label{VvsSigma.fig}Selection of the
extended cluster samples in the $V - \sigma_{\rm G}$ plane.  All sources
with $\sigma_{\rm G} \ge 1.25$ pixels are shown as filled dots; smaller
sources are plotted as open circles.  The selection limits at
$\sigma_{\rm G} = 1.25$ pixels and the 50\% completeness limits are
shown as the dashed lines.  Our final, verified samples are represented
by the sources in the upper left section of the $V - \sigma_{\rm G}$
plane.}

\figcaption[RdeGrijs.fig11.ps]{\label{NIRstars.fig}Near-IR
color-magnitude diagrams for the stellar background in the disk of M82. 
The thick lines represent theoretical isochrones for ages $t{\rm (yr)} =
10$, 30, and 100 Myr (top to bottom; Girardi et al.\ 2000, Salasnich et
al.\ 2000).  Typical photometric uncertainties and the direction of the
reddening vector are indicated.  For comparison, we have also included K
and M supergiants in the LMC (open circles; Oestreicher et al.\ 1997).}

\figcaption[RdeGrijs.fig12.ps]{\label{colcol.fig}Optical color-color
diagrams for the M82 B1 and B2 cluster samples.  Solid dots: $V \le
21.0$; open circles: $21.0 < V \le 22.5$.  The heavy solid lines are the
locations of unextincted, single-generation models of different ages by
BC96.  The thin lines indicate the effects of reddening on these
models.}

\figcaption[RdeGrijs.fig13.ps]{\label{cluscols.fig}$(B-V)$ and $(V-I)$
color distributions of the candidate clusters in the M82 B regions and
those of comparison samples of luminous {young} star cluster systems
taken from the literature.  Internal extinction corrections were applied
to the individual M82 B sources; the other color histograms were
corrected only for foreground Galactic extinction (Burstein \& Heiles
1984).  {\it (a)} \& {\it (f)} and {\it (b)} \& {\it (g)} M82 B1 and B2,
respectively (this work).  The shaded histograms represent the samples
with well-determined colors; the open histograms contain sources for
which only upper or lower limits could be obtained; {\it (c)} and {\it
(i)} NGC 1275 (Richer et al.  1993); {\it (d)} and {\it (k)} NGC 4038/39
(Whitmore et al.  1999) -- open: young clusters ($< 30$ Myr); shaded:
intermediate-age clusters ($0.25 - 1.0$ Gyr); {\it (e)} and {\it (l)}
NGC 7252 (Miller et al.  1997); {\it (h)} M82 A (O'Connell et al. 
1995); {\it (j)} NGC 3921 (Schweizer et al.  1996).}

\figcaption[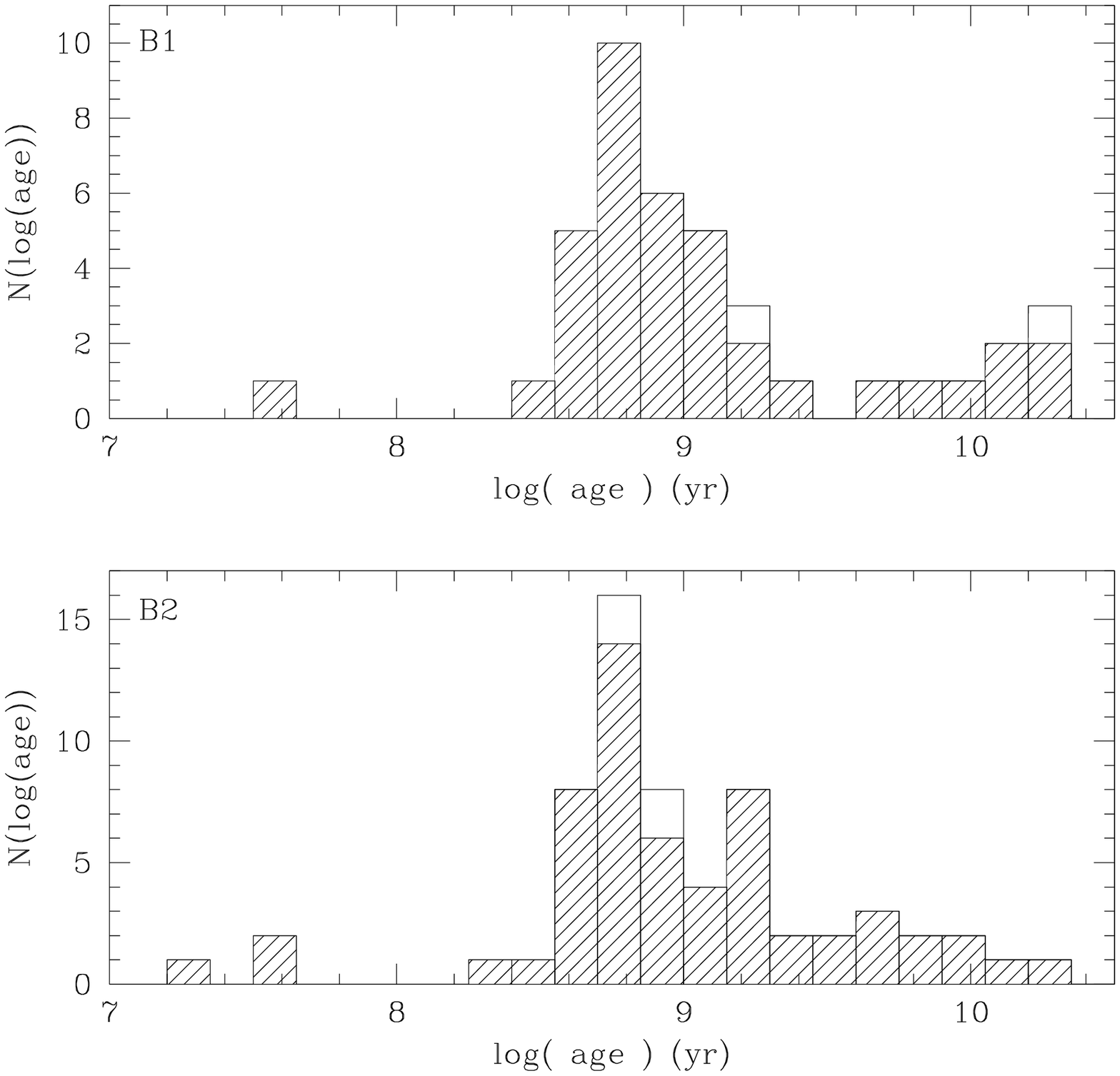]{\label{ages.fig}Age distribution of the
star clusters in M82 B1 and B2.  Shaded histogram: well-determined ages;
open histogram: upper or lower limits only, see Tables
\ref{derivedB1.tab} and \ref{derivedB2.tab}.}

\figcaption[RdeGrijs.fig15.ps]{\label{absv.fig} Cluster luminosity
functions of the M82 B regions, compared to the Galactic globular
cluster population and to recently published luminous {young} star
cluster systems in other nearby galaxies, based on {\sl HST}
observations.  The M82 CLFs have been corrected for internal extinction;
the other CLFs were corrected only for Galactic foreground extinction
(Burstein \& Heiles 1984).  Extinction corrections towards Milky Way
globular clusters were adopted from Harris (1996).  Where available we
have indicated the completeness limits by the dashed lines.  If no
independent distance estimate was available, distance moduli were
calculated for $H_0 = 50$ km s$^{-1}$ Mpc$^{-1}$.  {\it (a)} and {\it
(b)} M82 B1 and B2 (this paper; only sources with $V \le 22.5$ and
$\sigma_{\rm G} \ge 1.25$ pixels are included, the shaded CLFs represent
the samples with well-determined magnitudes; the open histograms contain
sources for which only lower limits could be determined); {\it (c)}
Milky Way globular clusters (Harris 1996); {\it (d)} M82 A (O'Connell et
al.  1995); {\it (e)} NGC 1275 (Carlson et al.  1998); {\it (f)} NGC
3597 (Holtzman et al.  1996); {\it (g)} NGC 3921 (Schweizer et al. 
1996); {\it (h)} NGC 4038/39 (Whitmore et al.  1999) -- open CLF: young
clusters ($< 30$ Myr); shaded CLF: intermediate-age clusters ($0.25 -
1.0$ Gyr); {\it (i)} and {\it (j)} NGC 7252 (Whitmore et al.  1993;
Miller et al.  1997), inner ($r < 6''$) and outer ($r > 6''$) sample,
respectively.}

\figcaption[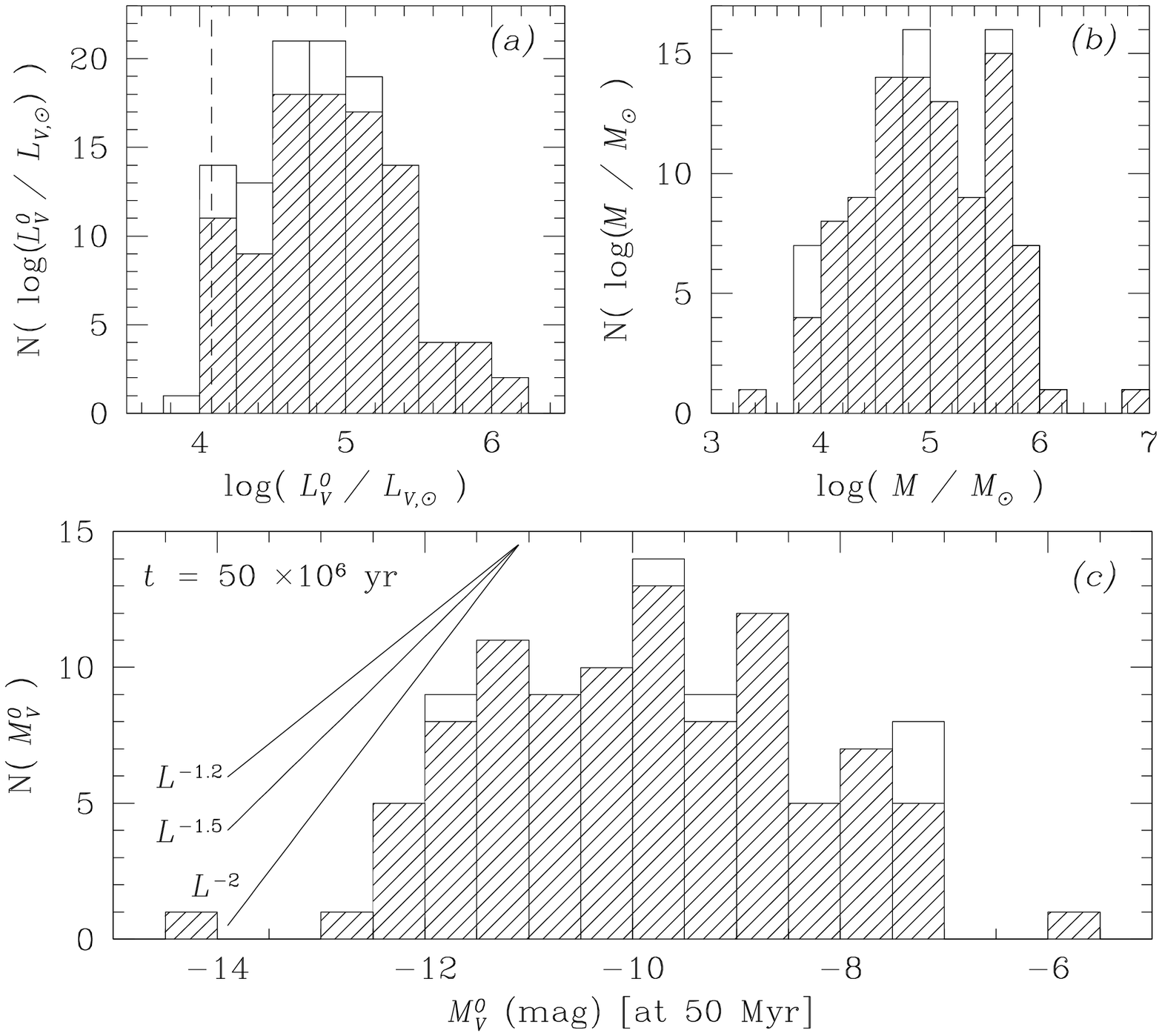]{\label{lumdist.fig}{\it (a)} Combined
{\it V}-band CLF for M82 B1 and B2, in units of {\it V}-band solar
luminosities and corrected for internal extinction.  The 50\%
completeness limit is indicated by the vertical dashed line; {\it (b)}
M82 B cluster mass distribution; {\it (c)} CLF for the M82 B cluster
sample corrected individually to a fiducial age of 50 Myr.  All panels:
shaded histograms -- well-determined source magnitudes; open histograms
-- lower limits only.}

\figcaption[RdeGrijs.fig17.ps]{\label{bright.fig}Luminosity profiles (in
{\it V}) for the 5 brightest candidate star clusters in both B1 and B2. 
Columns 1 and 3 show the actual data points, superposed with the
best-fitting Gaussian (short-dashed lines), modified Hubble ($\gamma =
2.0$, long-dashed lines), and $\gamma = 2.6$ profiles (solid lines;
Elson et al.\ 1987).  In columns 2 and 4, the flux ratios of data points
to model fit are presented: the dots represent the Gaussian fits, the
crosses a modified Hubble profile.}

\newpage
\begin{table}
\caption[ ]{\label{background.tab}Background surface brightnesses and
statistics of the M82 B fields}
\begin{center}
\begin{tabular}{cccccccc}
\hline
\hline
Region & Filter  & Mode & $\sigma$ & Region & Filter  & Mode & $\sigma$ \\
       &         & \multicolumn{2}{c}{(mag arcsec$^{-2}$)} & & & 
 \multicolumn{2}{c}{(mag arcsec$^{-2}$)} \\
\hline
B1     & {\it B} & 19.58 & 0.76 & B2     & {\it B} & 19.58 & 0.57 \\
       & {\it V} & 18.68 & 0.66 &        & {\it V} & 18.75 & 0.52 \\
       & {\it I} & 17.35 & 0.55 &        & {\it I} & 17.34 & 0.46 \\
       & {\it J} & 16.25 & 0.54 &        & {\it J} & 15.96 & 0.74 \\
       & {\it H} & 15.64 & 0.53 &        & {\it H} & 15.26 & 0.76 \\
\hline
\end{tabular}
\end{center}
\end{table}

\newpage

\begin{table}
\caption[ ]{\label{srcB1.tab}Observed properties of the cluster sample
in M82 B1}
{\scriptsize
\begin{center}
\begin{tabular}{rrcccccccc}
\hline
\hline
\multicolumn{1}{c}{\#} & \multicolumn{1}{c}{$\Delta$ RA$^1$} &
\multicolumn{1}{c}{$\Delta$ Dec$^1$} & \multicolumn{1}{c}{$\sigma_{\rm G}$} &
\multicolumn{1}{c}{$B$} & \multicolumn{1}{c}{$V$} &
\multicolumn{1}{c}{$I$} & \multicolumn{1}{c}{$J$} &
\multicolumn{1}{c}{$H$} & Notes \\
\cline{5-9}
& (sec) & (arcsec) & (arcsec) & \multicolumn{5}{c}{(mag)} \\
\hline
 1 &--1.33 & 1.55 & 0.07 & $22.86 \pm 0.08$ & $21.59 \pm 0.03$ & $19.89 \pm 0.02$ & $18.70 \pm 0.12$ & $18.14 \pm 0.12$\\
 2 &--0.01 & 1.47 & 0.13 & $21.99 \pm 0.07$ & $21.18 \pm 0.06$ & $20.15 \pm 0.05$ & $20.32 \pm 0.21$ & $>19.42        $\\
 3 &  0.28 & 1.44 & 0.10 & $>22.97        $ & $22.04 \pm 0.11$ & $19.61 \pm 0.06$ & $18.25 \pm 0.10$ & $>17.64        $\\
 4 &--1.16 & 4.71 & 0.13 & $21.91 \pm 0.07$ & $20.91 \pm 0.05$ & $19.16 \pm 0.03$ & $17.76 \pm 0.07$ & $17.11 \pm 0.07$\\
 5 &  1.62 &-0.26 & 0.12 & $21.90 \pm 0.05$ & $20.69 \pm 0.03$ & $19.18 \pm 0.02$ & $18.12 \pm 0.08$ & $17.70 \pm 0.08$\\
 6 &  0.57 & 2.49 & 0.14 & $20.47 \pm 0.06$ & $19.71 \pm 0.04$ & $18.32 \pm 0.02$ & $17.18 \pm 0.06$ & $16.65 \pm 0.05$\\
 7 &  0.75 & 5.70 & 0.09 & $19.96 \pm 0.04$ & $19.17 \pm 0.03$ & $18.08 \pm 0.02$ & $17.20 \pm 0.06$ & $16.75 \pm 0.05$\\
 8 &  0.86 & 6.44 & 0.13 & $19.61 \pm 0.03$ & $19.19 \pm 0.02$ & $18.46 \pm 0.03$ & $17.75 \pm 0.07$ & $17.42 \pm 0.08$\\
 9 &  1.18 & 6.95 & 0.09 & $22.34 \pm 0.12$ & $21.81 \pm 0.11$ & $21.14 \pm 0.18$ & $19.44 \pm 0.18$ & $>18.73        $\\
10 &  0.80 & 8.10 & 0.06 & $22.46 \pm 0.11$ & $21.89 \pm 0.08$ & $20.93 \pm 0.09$ & $21.34 \pm 0.63$ & $>19.87        $ & $^2$\\
11 &  0.84 & 8.38 & 0.11 & $19.71 \pm 0.03$ & $19.15 \pm 0.03$ & $18.30 \pm 0.03$ & $17.55 \pm 0.07$ & $17.24 \pm 0.08$\\
12 &  0.03 & 0.55 & 0.09 & $21.49 \pm 0.08$ & $20.76 \pm 0.08$ & $19.79 \pm 0.07$ & $18.77 \pm 0.13$ & $17.92 \pm 0.11$\\
13 &  1.11 & 8.95 & 0.13 & $22.00 \pm 0.07$ & $21.25 \pm 0.07$ & $>20.89        $ & $>20.62        $ & $>19.75        $ & $^2$\\
14 &  0.41 & 0.73 & 0.14 & $21.49 \pm 0.08$ & $20.59 \pm 0.05$ & $19.19 \pm 0.03$ & $17.91 \pm 0.10$ & $17.29 \pm 0.09$\\
15 &  0.33 & 0.87 & 0.16 & $>21.31        $ & $19.83 \pm 0.04$ & $18.12 \pm 0.02$ & $16.99 \pm 0.05$ & $16.50 \pm 0.05$\\
16 &  0.24 & 1.65 & 0.07 & $23.48 \pm 0.24$ & $22.08 \pm 0.10$ & $20.17 \pm 0.03$ & $19.07 \pm 0.15$ & $18.17 \pm 0.13$\\
17 &  2.16 & 8.33 & 0.08 & $22.67 \pm 0.08$ & $21.88 \pm 0.06$ & $20.75 \pm 0.09$ & $18.79 \pm 0.14$ & $>17.83        $\\
18 &  2.47 & 8.10 & 0.11 & $19.65 \pm 0.02$ & $19.01 \pm 0.02$ & $18.03 \pm 0.02$ & $17.18 \pm 0.06$ & $16.59 \pm 0.05$\\
19 &  3.28 & 6.70 & 0.08 & $>22.55        $ & $21.15 \pm 0.06$ & $19.31 \pm 0.02$ & $18.35 \pm 0.09$ & $>17.80        $\\
20 &  2.22 & 9.04 & 0.12 & $19.87 \pm 0.03$ & $19.12 \pm 0.02$ & $18.11 \pm 0.03$ & $17.68 \pm 0.08$ & $17.36 \pm 0.09$\\
21 &--0.20 & 4.31 & 0.14 & $21.75 \pm 0.04$ & $21.24 \pm 0.05$ & $>20.25        $ & $19.75 \pm 0.21$ & $18.76 \pm 0.17$\\
22 &  1.41 & 2.26 & 0.10 & $20.47 \pm 0.03$ & $19.80 \pm 0.03$ & $18.70 \pm 0.03$ & $17.52 \pm 0.07$ & $16.82 \pm 0.07$\\
23 &  1.56 & 3.07 & 0.08 & $>22.07        $ & $21.32 \pm 0.07$ & $19.67 \pm 0.02$ & $18.70 \pm 0.11$ & $18.26 \pm 0.12$\\
24 &  1.25 & 4.06 & 0.12 & $20.39 \pm 0.03$ & $19.79 \pm 0.04$ & $18.85 \pm 0.04$ & $17.98 \pm 0.09$ & $17.33 \pm 0.08$\\
25 &  0.08 & 6.32 & 0.15 & $21.57 \pm 0.05$ & $20.70 \pm 0.03$ & $19.47 \pm 0.03$ & $18.83 \pm 0.12$ & $18.46 \pm 0.13$\\
26 &  2.49 & 2.18 & 0.11 & $21.51 \pm 0.08$ & $21.18 \pm 0.08$ & $20.71 \pm 0.11$ & $19.70 \pm 0.21$ & $>18.35        $\\
27 &  2.94 & 2.44 & 0.12 & $21.66 \pm 0.04$ & $21.16 \pm 0.04$ & $20.48 \pm 0.07$ & $20.51 \pm 0.27$ & $>20.11        $\\
28 &  3.34 & 2.22 & 0.11 & $18.56 \pm 0.01$ & $17.93 \pm 0.01$ & $16.97 \pm 0.01$ & $16.15 \pm 0.03$ & $15.72 \pm 0.03$\\
29 &--0.07 & 8.38 & 0.12 & $22.00 \pm 0.09$ & $21.31 \pm 0.06$ & $20.17 \pm 0.06$ & $19.14 \pm 0.14$ & $18.42 \pm 0.12$\\
30 &  3.33 & 3.38 & 0.06 & $20.63 \pm 0.02$ & $20.09 \pm 0.02$ & $19.27 \pm 0.03$ & $18.65 \pm 0.12$ & $18.19 \pm 0.11$\\
31 &  3.75 & 2.76 & 0.14 & $>22.77        $ & $21.63 \pm 0.08$ & $20.24 \pm 0.06$ & $19.57 \pm 0.18$ & $19.18 \pm 0.19$ & $^2$\\
32 &  2.51 & 5.33 & 0.07 & $21.95 \pm 0.07$ & $21.24 \pm 0.05$ & $20.26 \pm 0.05$ & $19.80 \pm 0.22$ & $19.63 \pm 0.28$\\
33 &  1.67 & 7.17 & 0.07 & $23.44 \pm 0.39$ & $22.46 \pm 0.22$ & $21.28 \pm 0.11$ & $20.58 \pm 0.29$ & $20.79 \pm 0.55$ & $^2$\\
34 &  2.68 & 6.30 & 0.12 & $21.89 \pm 0.13$ & $21.46 \pm 0.08$ & $20.42 \pm 0.06$ & $21.14 \pm 1.10$ & $21.32 \pm 2.04$\\
35 &  0.84 & 9.80 & 0.20 & $21.41 \pm 0.05$ & $20.63 \pm 0.05$ & $19.43 \pm 0.04$ & $19.17 \pm 0.15$ & $19.05 \pm 0.19$\\
36 &  3.76 & 4.81 & 0.14 & $22.46 \pm 0.14$ & $21.19 \pm 0.08$ & $19.56 \pm 0.04$ & $18.29 \pm 0.10$ & $17.47 \pm 0.08$\\
37 &  2.11 & 8.34 & 0.10 & $20.49 \pm 0.04$ & $19.51 \pm 0.02$ & $18.17 \pm 0.02$ & $17.25 \pm 0.06$ & $16.88 \pm 0.06$\\
38 &  3.48 & 0.34 & 0.11 & $21.92 \pm 0.08$ & $21.29 \pm 0.06$ & $20.28 \pm 0.07$ & $19.03 \pm 0.15$ & $18.44 \pm 0.14$\\
39 &  5.96 & 6.34 & 0.10 & $23.41 \pm 0.09$ & $22.48 \pm 0.07$ & $21.22 \pm 0.09$ & $22.05 \pm 0.75$ & $>19.17        $\\
40 &  3.46 & 2.39 & 0.07 & $22.94 \pm 0.07$ & $22.48 \pm 0.07$ & $21.67 \pm 0.11$ & $20.55 \pm 0.28$ & $20.16 \pm 0.30$\\
41 &  3.81 & 2.32 & 0.18 & $21.32 \pm 0.06$ & $20.64 \pm 0.04$ & $19.52 \pm 0.03$ & $18.47 \pm 0.10$ & $17.78 \pm 0.09$\\
42 &  4.92 & 1.45 & 0.17 & $>22.61        $ & $21.35 \pm 0.03$ & $20.13 \pm 0.04$ & $19.32 \pm 0.15$ & $18.70 \pm 0.13$\\
43 &  1.19 & 0.62 & 0.11 & $23.35 \pm 0.09$ & $22.38 \pm 0.05$ & $21.14 \pm 0.05$ & $19.49 \pm 0.23$ & $20.34 \pm 0.25$\\
\hline
\end{tabular}
\end{center}
$^1$ -- Offsets w.r.t. J2000 coordinates RA = 09:56:00.0; Dec = 69:41:00.0\\
$^2$ -- Source identifications marginal
}
\end{table}

\newpage

\begin{table}
\caption[ ]{\label{srcB2.tab}Observed properties of the cluster sample
in M82 B2}
{\scriptsize
\begin{center}
\begin{tabular}{rrcccccccc}
\hline
\hline
\multicolumn{1}{c}{\#} & \multicolumn{1}{c}{$\Delta$ RA$^1$} &
\multicolumn{1}{c}{$\Delta$ Dec$^1$} & \multicolumn{1}{c}{$\sigma_G$} &
\multicolumn{1}{c}{$B$} & \multicolumn{1}{c}{$V$} &
\multicolumn{1}{c}{$I$} & \multicolumn{1}{c}{$J$} &
\multicolumn{1}{c}{$H$} & Notes \\
\cline{5-9}
& (sec) & (arcsec) & (arcsec) & \multicolumn{5}{c}{(mag)} \\
\hline
 1 & 58.35 & 47.94 & 0.15 & $23.07 \pm 0.17$ & $22.01 \pm 0.09$ & $20.86 \pm 0.09$ & $19.59 \pm 0.16$ & $19.53 \pm 0.20$\\
 2 & 54.76 & 55.37 & 0.09 & $>22.32        $ & $20.23 \pm 0.03$ & $17.21 \pm 0.01$ & $14.75 \pm 0.02$ & $>13.74        $\\
 3 & 59.18 & 49.03 & 0.11 & $>23.01        $ & $21.37 \pm 0.03$ & $19.08 \pm 0.02$ & $\dots         $ & $\dots         $\\
 4 & 54.78 & 57.31 & 0.16 & $21.01 \pm 0.02$ & $20.12 \pm 0.02$ & $18.89 \pm 0.04$ & $16.93 \pm 0.07$ & $>15.76        $\\
 5 & 55.98 & 56.89 & 0.24 & $20.41 \pm 0.04$ & $19.77 \pm 0.05$ & $18.61 \pm 0.10$ & $18.15 \pm 0.22$ & $17.98 \pm 0.37$\\
 6 & 59.20 & 51.77 & 0.22 & $>23.12        $ & $21.33 \pm 0.05$ & $19.48 \pm 0.03$ & $14.43 \pm 0.01$ & $>17.39        $ & $^2$\\
 7 & 55.76 & 58.66 & 0.15 & $21.73 \pm 0.07$ & $20.58 \pm 0.06$ & $18.84 \pm 0.06$ & $>18.40        $ & $>19.04        $\\
 8 & 56.08 & 58.16 & 0.10 & $23.69 \pm 0.11$ & $22.00 \pm 0.06$ & $19.44 \pm 0.04$ & $17.84 \pm 0.08$ & $>17.21        $\\
 9 & 56.36 & 57.82 & 0.08 & $22.80 \pm 0.10$ & $22.31 \pm 0.20$ & $20.53 \pm 0.14$ & $20.32 \pm 0.31$ & $>17.70        $ & $^2$\\
10 & 55.92 & 58.72 & 0.07 & $>23.23        $ & $22.10 \pm 0.04$ & $20.33 \pm 0.09$ & $18.91 \pm 0.16$ & $18.36 \pm 0.21$\\
11 & 56.19 & 58.33 & 0.08 & $>23.48        $ & $22.46 \pm 0.11$ & $20.50 \pm 0.09$ & $19.26 \pm 0.17$ & $18.69 \pm 0.17$ & $^2$\\
12 & 54.54 & 61.69 & 0.10 & $19.04 \pm 0.02$ & $17.90 \pm 0.01$ & $16.40 \pm 0.01$ & $15.17 \pm 0.03$ & $14.58 \pm 0.03$\\
13 & 53.44 & 63.70 & 0.13 & $21.96 \pm 0.05$ & $20.84 \pm 0.03$ & $19.38 \pm 0.03$ & $18.41 \pm 0.10$ & $18.00 \pm 0.09$\\
14 & 56.37 & 58.45 & 0.14 & $23.77 \pm 0.18$ & $22.45 \pm 0.04$ & $20.58 \pm 0.10$ & $18.30 \pm 0.14$ & $>16.76        $ & $^2$\\
15 & 57.14 & 57.65 & 0.15 & $22.35 \pm 0.05$ & $20.79 \pm 0.03$ & $18.38 \pm 0.03$ & $16.88 \pm 0.07$ & $16.08 \pm 0.06$\\
16 & 55.85 & 60.16 & 0.10 & $>23.44        $ & $22.32 \pm 0.06$ & $19.71 \pm 0.03$ & $17.90 \pm 0.09$ & $>17.11        $ \\
17 & 55.25 & 61.31 & 0.24 & $21.88 \pm 0.07$ & $20.76 \pm 0.06$ & $19.22 \pm 0.05$ & $18.20 \pm 0.10$ & $17.56 \pm 0.09$\\
18 & 57.15 & 58.30 & 0.08 & $23.52 \pm 0.15$ & $22.17 \pm 0.10$ & $19.58 \pm 0.07$ & $18.32 \pm 0.15$ & $>17.30        $ \\
19 & 56.55 & 60.17 & 0.14 & $>23.22        $ & $20.68 \pm 0.04$ & $17.80 \pm 0.02$ & $15.57 \pm 0.03$ & $>14.59        $ \\
20 & 54.34 & 65.94 & 0.09 & $>22.88        $ & $21.52 \pm 0.04$ & $19.80 \pm 0.03$ & $18.84 \pm 0.12$ & $18.50 \pm 0.12$\\
21 & 57.45 & 60.53 & 0.09 & $22.55 \pm 0.07$ & $21.30 \pm 0.06$ & $18.73 \pm 0.04$ & $16.55 \pm 0.05$ & $>15.59        $ & $^2$\\
22 & 56.62 & 62.28 & 0.07 & $23.64 \pm 0.16$ & $22.42 \pm 0.08$ & $20.91 \pm 0.10$ & $20.85 \pm 0.32$ & $>19.01        $ \\
23 & 57.90 & 59.97 & 0.08 & $23.62 \pm 0.23$ & $22.46 \pm 0.21$ & $20.76 \pm 0.19$ & $19.54 \pm 0.24$ & $18.84 \pm 0.21$ & $^2$\\
24 & 57.01 & 62.20 & 0.11 & $>22.27        $ & $21.24 \pm 0.08$ & $19.41 \pm 0.04$ & $18.14 \pm 0.09$ & $17.63 \pm 0.09$ & $^2$\\
25 & 56.06 & 64.02 & 0.17 & $20.97 \pm 0.06$ & $20.32 \pm 0.07$ & $19.17 \pm 0.07$ & $18.10 \pm 0.10$ & $17.39 \pm 0.08$\\
26 & 55.59 & 65.91 & 0.13 & $19.11 \pm 0.01$ & $18.57 \pm 0.01$ & $17.68 \pm 0.01$ & $16.96 \pm 0.05$ & $16.55 \pm 0.05$\\
27 & 56.94 & 63.62 & 0.06 & $22.77 \pm 0.15$ & $22.44 \pm 0.28$ & $21.20 \pm 0.24$ & $19.94 \pm 0.25$ & $19.21 \pm 0.20$\\
28 & 58.09 & 62.27 & 0.08 & $22.43 \pm 0.11$ & $21.78 \pm 0.10$ & $20.29 \pm 0.08$ & $19.40 \pm 0.22$ & $18.74 \pm 0.19$ & $^2$\\
29 & 55.11 & 67.74 & 0.15 & $22.82 \pm 0.10$ & $22.31 \pm 0.07$ & $21.33 \pm 0.11$ & $21.10 \pm 0.43$ & $20.46 \pm 0.44$\\
30 & 55.55 & 66.99 & 0.13 & $21.77 \pm 0.07$ & $20.87 \pm 0.04$ & $19.43 \pm 0.03$ & $18.50 \pm 0.12$ & $18.05 \pm 0.13$ & $^2$\\
31 & 58.14 & 62.52 & 0.08 & $22.07 \pm 0.08$ & $21.50 \pm 0.09$ & $20.46 \pm 0.10$ & $20.12 \pm 0.21$ & $>18.58        $ \\
32 & 58.71 & 61.61 & 0.07 & $22.65 \pm 0.09$ & $21.48 \pm 0.04$ & $19.78 \pm 0.03$ & $18.63 \pm 0.11$ & $18.24 \pm 0.12$\\
33 & 58.06 & 62.92 & 0.07 & $23.19 \pm 0.17$ & $22.47 \pm 0.16$ & $21.29 \pm 0.18$ & $19.57 \pm 0.21$ & $>18.32        $ & $^2$\\
34 & 55.30 & 68.13 & 0.10 & $23.04 \pm 0.07$ & $22.48 \pm 0.05$ & $21.40 \pm 0.07$ & $20.31 \pm 0.25$ & $19.67 \pm 0.24$\\
35 & 57.88 & 63.63 & 0.09 & $>22.86        $ & $22.04 \pm 0.15$ & $20.38 \pm 0.10$ & $19.02 \pm 0.15$ & $18.28 \pm 0.13$ & $^2$\\
36 & 54.01 & 71.07 & 0.19 & $20.45 \pm 0.02$ & $19.72 \pm 0.01$ & $18.82 \pm 0.02$ & $17.99 \pm 0.08$ & $17.52 \pm 0.08$\\
37 & 57.69 & 64.55 & 0.13 & $20.55 \pm 0.06$ & $19.81 \pm 0.03$ & $18.15 \pm 0.04$ & $16.72 \pm 0.05$ & $16.12 \pm 0.05$\\
38 & 55.14 & 69.71 & 0.14 & $21.11 \pm 0.04$ & $20.66 \pm 0.04$ & $20.09 \pm 0.06$ & $19.55 \pm 0.19$ & $18.86 \pm 0.17$\\
39 & 58.01 & 65.15 & 0.07 & $21.69 \pm 0.18$ & $21.23 \pm 0.17$ & $20.66 \pm 0.22$ & $19.58 \pm 0.23$ & $>18.07        $ & $^2$\\
40 & 55.43 & 69.94 & 0.13 & $19.70 \pm 0.01$ & $19.31 \pm 0.01$ & $18.56 \pm 0.02$ & $17.88 \pm 0.08$ & $17.45 \pm 0.08$\\
41 & 56.98 & 67.39 & 0.11 & $19.39 \pm 0.02$ & $18.56 \pm 0.02$ & $17.38 \pm 0.02$ & $16.19 \pm 0.04$ & $15.88 \pm 0.04$\\
42 & 59.80 & 62.46 & 0.18 & $23.39 \pm 0.28$ & $21.48 \pm 0.10$ & $>19.18        $ & $17.98 \pm 0.09$ & $>17.01        $ & $^2$\\
43 & 55.92 & 70.09 & 0.10 & $21.35 \pm 0.03$ & $20.91 \pm 0.02$ & $20.34 \pm 0.05$ & $20.45 \pm 0.26$ & $>18.74        $\\
44 & 56.22 & 69.60 & 0.15 & $21.38 \pm 0.05$ & $20.91 \pm 0.04$ & $20.09 \pm 0.05$ & $20.38 \pm 0.23$ & $>18.36        $\\
45 & 58.20 & 67.38 & 0.09 & $22.85 \pm 0.10$ & $21.68 \pm 0.05$ & $19.95 \pm 0.03$ & $19.27 \pm 0.15$ & $18.75 \pm 0.15$\\
46 & 54.88 & 73.60 & 0.08 & $22.30 \pm 0.05$ & $22.17 \pm 0.04$ & $>21.55        $ & $>19.97        $ & $>19.43        $\\
47 & 56.65 & 70.80 & 0.10 & $22.75 \pm 0.07$ & $21.99 \pm 0.05$ & $21.06 \pm 0.06$ & $20.92 \pm 0.27$ & $>20.97        $\\
48 & 54.70 & 74.99 & 0.14 & $22.39 \pm 0.05$ & $21.84 \pm 0.05$ & $20.83 \pm 0.05$ & $20.04 \pm 0.22$ & $19.38 \pm 0.20$\\
49 & 58.35 & 68.76 & 0.11 & $20.97 \pm 0.05$ & $19.96 \pm 0.04$ & $18.65 \pm 0.04$ & $17.66 \pm 0.08$ & $17.40 \pm 0.07$\\
50 & 57.92 & 70.35 & 0.09 & $22.09 \pm 0.09$ & $21.46 \pm 0.08$ & $20.12 \pm 0.05$ & $19.14 \pm 0.15$ & $18.61 \pm 0.14$\\
\multicolumn{5}{l}{(continued on next page)}\\
\hline
\end{tabular}
\end{center}
}
\end{table}

\newpage

\addtocounter{table}{-1}
\begin{table}
\caption[ ]{(continued)}
{\scriptsize
\begin{center}
\begin{tabular}{rrcccccccc}
\hline
\hline
\multicolumn{1}{c}{\#} & \multicolumn{1}{c}{$\Delta$ RA$^1$} &
\multicolumn{1}{c}{$\Delta$ Dec$^1$} & \multicolumn{1}{c}{$\sigma_G$} &
\multicolumn{1}{c}{$B$} & \multicolumn{1}{c}{$V$} &
\multicolumn{1}{c}{$I$} & \multicolumn{1}{c}{$J$} &
\multicolumn{1}{c}{$H$} & Notes \\
\cline{5-9}
& (sec) & (arcsec) & (arcsec) & \multicolumn{5}{c}{(mag)} \\
\hline
\multicolumn{5}{l}{(continued from previous page)}\\
51 & 60.41 & 67.54 & 0.08 & $22.66 \pm 0.11$ & $22.17 \pm 0.11$ & $21.64 \pm 0.20$ & $>20.12        $ & $>19.28        $\\
52 & 57.88 & 72.53 & 0.15 & $21.58 \pm 0.06$ & $20.93 \pm 0.05$ & $20.16 \pm 0.07$ & $19.99 \pm 0.22$ & $20.09 \pm 0.38$ & $^2$\\
53 & 58.86 & 70.80 & 0.10 & $21.18 \pm 0.05$ & $20.38 \pm 0.04$ & $19.60 \pm 0.04$ & $19.08 \pm 0.15$ & $18.52 \pm 0.15$\\
54 & 58.06 & 72.79 & 0.13 & $21.24 \pm 0.04$ & $20.91 \pm 0.05$ & $20.34 \pm 0.08$ & $19.73 \pm 0.20$ & $19.37 \pm 0.22$\\
55 & 55.58 & 77.50 & 0.11 & $22.57 \pm 0.06$ & $22.15 \pm 0.04$ & $21.64 \pm 0.07$ & $20.61 \pm 0.32$ & $>19.58        $\\
56 & 60.05 & 70.62 & 0.12 & $21.52 \pm 0.09$ & $20.77 \pm 0.07$ & $19.81 \pm 0.08$ & $18.69 \pm 0.13$ & $17.86 \pm 0.12$\\
57 & 60.42 & 70.80 & 0.16 & $21.61 \pm 0.08$ & $20.59 \pm 0.05$ & $19.15 \pm 0.03$ & $17.49 \pm 0.06$ & $17.07 \pm 0.06$\\
58 & 60.35 & 70.94 & 0.15 & $>21.15        $ & $19.81 \pm 0.04$ & $18.14 \pm 0.02$ & $17.13 \pm 0.06$ & $16.67 \pm 0.06$\\
59 & 60.26 & 71.73 & 0.07 & $23.19 \pm 0.14$ & $22.15 \pm 0.10$ & $20.27 \pm 0.04$ & $19.33 \pm 0.17$ & $18.78 \pm 0.16$\\
60 & 58.50 & 76.27 & 0.17 & $22.15 \pm 0.06$ & $21.56 \pm 0.06$ & $20.58 \pm 0.05$ & $20.00 \pm 0.21$ & $19.35 \pm 0.20$ & $^2$\\
61 & 59.83 & 74.32 & 0.13 & $22.45 \pm 0.06$ & $21.97 \pm 0.06$ & $>21.13        $ & $20.79 \pm 0.30$ & $20.62 \pm 0.43$ & $^2$\\
62 & 58.77 & 77.33 & 0.09 & $22.88 \pm 0.07$ & $22.30 \pm 0.05$ & $21.54 \pm 0.07$ & $21.10 \pm 0.32$ & $21.06 \pm 0.38$\\
63 & 57.83 & 79.12 & 0.17 & $21.47 \pm 0.03$ & $20.93 \pm 0.02$ & $20.00 \pm 0.04$ & $19.24 \pm 0.15$ & $18.89 \pm 0.16$\\
64 & 55.66 & 84.39 & 0.08 & $23.14 \pm 0.07$ & $22.36 \pm 0.04$ & $21.42 \pm 0.06$ & $20.87 \pm 0.35$ & $19.90 \pm 0.27$\\
65 & 60.11 & 76.41 & 0.16 & $21.57 \pm 0.05$ & $20.70 \pm 0.03$ & $19.46 \pm 0.03$ & $19.04 \pm 0.14$ & $18.72 \pm 0.17$\\
66 & 60.01 & 76.87 & 0.08 & $23.18 \pm 0.08$ & $22.66 \pm 0.10$ & $21.89 \pm 0.09$ & $20.77 \pm 0.31$ & $19.94 \pm 0.26$\\
67 & 57.72 & 82.21 & 0.23 & $21.56 \pm 0.04$ & $20.64 \pm 0.02$ & $19.21 \pm 0.02$ & $18.06 \pm 0.09$ & $17.38 \pm 0.08$\\
68 & 59.95 & 78.41 & 0.13 & $22.43 \pm 0.08$ & $21.63 \pm 0.05$ & $20.56 \pm 0.04$ & $19.58 \pm 0.18$ & $19.03 \pm 0.17$\\
69 & 55.50 & 87.07 & 0.09 & $22.76 \pm 0.06$ & $22.19 \pm 0.04$ & $21.49 \pm 0.07$ & $20.71 \pm 0.29$ & $20.23 \pm 0.29$\\
70 & 58.21 & 83.36 & 0.11 & $20.40 \pm 0.01$ & $19.93 \pm 0.01$ & $19.20 \pm 0.02$ & $18.91 \pm 0.12$ & $>18.09        $\\
\hline
\end{tabular}
\end{center}
$^1$ -- Offsets w.r.t. J2000 coordinates RA = 09:55:00.0; Dec = 69:40:00.0\\
$^2$ -- Source identifications marginal
}
\end{table}

\newpage

\begin{table}
\caption[ ]{\label{derivedB1.tab}Derived properties of the cluster sample
in M82 B1}
{\scriptsize
\begin{center}
\begin{tabular}{rrrrrrrr}
\hline
\hline
\multicolumn{1}{c}{\#} & \multicolumn{1}{c}{$A_V$} &
\multicolumn{1}{c}{$M_V^0$} & \multicolumn{1}{c}{$(B-V)_0$} &
\multicolumn{1}{c}{$(V-I)_0$} & \multicolumn{1}{c}{$(V-H)_0$} &
\multicolumn{1}{c}{$(J-H)_0$} & \multicolumn{1}{c}{log( Age )} \\
\cline{2-7}
& \multicolumn{6}{c}{(mag)} & \multicolumn{1}{c}{(yr)} \\
\hline
 1 & 0.95 & $ -7.16$ & $ 0.96$ & $ 1.21$ & $ 2.67$ & $ 0.46$ & $ 10.2$ \\
 2 & 0.11 & $ -6.73$ & $ 0.77$ & $ 0.97$ & $<1.67$ & $<0.89$ & $  9.3$ \\
 3 &$\dots$&$<-5.76$ & $\dots$ & $<2.43$ & $<4.40$ & $<0.61$ & $\dots$ \\
 4 & 1.81 & $ -8.70$ & $ 0.41$ & $ 0.81$ & $ 2.31$ & $ 0.46$ & $  8.8$ \\
 5 & 0.56 & $ -7.67$ & $ 1.03$ & $ 1.22$ & $ 2.53$ & $ 0.36$ & $ 10.1$ \\
 6 & 1.37 & $ -9.46$ & $ 0.32$ & $ 0.68$ & $ 1.93$ & $ 0.38$ & $  8.7$ \\
 7 & 0.38 & $ -9.01$ & $ 0.67$ & $ 0.89$ & $ 2.11$ & $ 0.41$ & $  9.0$ \\
 8 & 0.28 & $ -8.89$ & $ 0.33$ & $ 0.58$ & $ 1.54$ & $ 0.30$ & $  8.7$ \\
 9 & 0.00 & $ -5.99$ & $ 0.53$ & $ 0.67$ & $<3.08$ & $<0.71$ & $  8.9$ \\
10 & 0.56 & $ -6.47$ & $ 0.39$ & $ 0.67$ & $<1.56$ & $<1.41$ & $  8.8$ \\
11 & 0.27 & $ -8.92$ & $ 0.47$ & $ 0.71$ & $ 1.69$ & $ 0.28$ & $  8.8$ \\
12 & 0.18 & $ -7.22$ & $ 0.67$ & $ 0.88$ & $ 2.69$ & $ 0.83$ & $  9.0$ \\
13 &$\dots$&$<-6.55$ & $<0.75$ & $<0.36$ & $<1.50$ & $\dots$ & $ <9.2$ \\
14 & 1.06 & $ -8.27$ & $ 0.56$ & $ 0.85$ & $ 2.43$ & $ 0.51$ & $  8.9$ \\
15 &$\dots$&$<-7.97$ & $\dots$ & $<1.71$ & $<3.33$ & $<0.49$ & $\dots$ \\
16 & 1.23 & $ -6.95$ & $ 1.00$ & $ 1.27$ & $ 2.90$ & $ 0.77$ & $ 10.0$ \\
17 & 0.53 & $ -6.45$ & $ 0.62$ & $ 0.86$ & $<3.61$ & $<0.90$ & $  9.0$ \\
18 & 0.46 & $ -9.25$ & $ 0.49$ & $ 0.74$ & $ 2.04$ & $ 0.54$ & $  8.9$ \\
19 &$\dots$&$<-6.65$ & $\dots$ & $<1.84$ & $<3.35$ & $<0.55$ & $\dots$ \\
20 & 0.21 & $ -8.89$ & $ 0.68$ & $ 0.90$ & $ 1.59$ & $ 0.30$ & $  9.1$ \\
21 & 0.79 & $ -7.35$ & $ 0.25$ & $<0.58$ & $ 1.83$ & $ 0.91$ & $  8.5$ \\
22 & 0.76 & $ -8.76$ & $ 0.42$ & $ 0.71$ & $ 2.35$ & $ 0.62$ & $  8.8$ \\
23 & 2.15 & $ -8.63$ & $>0.05$ & $ 0.54$ & $ 1.29$ & $ 0.21$ & $  8.7$ \\
24 & 0.40 & $ -8.41$ & $ 0.47$ & $ 0.73$ & $ 2.13$ & $ 0.61$ & $  8.8$ \\
25 & 0.51 & $ -7.61$ & $ 0.70$ & $ 0.97$ & $ 1.82$ & $ 0.32$ & $  9.1$ \\
26 & 0.00 & $ -6.62$ & $ 0.33$ & $ 0.47$ & $<2.83$ & $<1.35$ & $  8.7$ \\
27 & 0.00 & $ -6.64$ & $ 0.50$ & $ 0.68$ & $<1.05$ & $<0.40$ & $  8.9$ \\
28 & 0.39 & $-10.26$ & $ 0.50$ & $ 0.76$ & $ 1.89$ & $ 0.39$ & $  8.9$ \\
29 & 0.76 & $ -7.25$ & $ 0.44$ & $ 0.75$ & $ 2.26$ & $ 0.64$ & $  8.8$ \\
30 & 0.26 & $ -7.97$ & $ 0.46$ & $ 0.69$ & $ 1.69$ & $ 0.43$ & $  8.8$ \\
31 & 0.30 & $ -6.47$ & $>1.04$ & $ 1.23$ & $ 2.20$ & $ 0.36$ & $ 10.3$ \\
32 & 0.30 & $ -6.86$ & $ 0.61$ & $ 0.82$ & $ 1.36$ & $ 0.14$ & $  9.0$ \\
33 & 0.15 & $ -5.49$ & $ 0.93$ & $ 1.10$ & $ 1.55$ & $-0.23$ & $  9.8$ \\
34 & 1.10 & $ -7.44$ & $ 0.07$ & $ 0.47$ & $-0.77$ & $-0.30$ & $  7.6$ \\
35 & 0.74 & $ -7.91$ & $ 0.54$ & $ 0.82$ & $ 0.97$ & $ 0.04$ & $  8.9$ \\
36 & 0.73 & $ -7.34$ & $ 1.03$ & $ 1.25$ & $ 3.12$ & $ 0.74$ & $ 10.1$ \\
37 & 0.57 & $ -8.86$ & $ 0.80$ & $ 1.04$ & $ 2.16$ & $ 0.31$ & $  9.3$ \\
38 & 0.51 & $ -7.02$ & $ 0.46$ & $ 0.75$ & $ 2.43$ & $ 0.54$ & $  8.8$ \\
39 & 0.46 & $ -5.78$ & $ 0.78$ & $ 1.02$ & $<2.93$ & $<2.83$ & $  9.3$ \\
40 & 0.40 & $ -5.72$ & $ 0.33$ & $ 0.60$ & $ 1.99$ & $ 0.35$ & $  8.7$ \\
41 & 0.75 & $ -7.91$ & $ 0.44$ & $ 0.73$ & $ 2.24$ & $ 0.61$ & $  8.8$ \\
42 &$\dots$&$<-6.45$ & $\dots$ & $<1.22$ & $<2.65$ & $<0.62$ & $<10.3$ \\
43 & 0.33 & $ -5.75$ & $ 0.86$ & $ 1.07$ & $ 1.77$ & $-0.89$ & $  9.7$ \\
\hline
\end{tabular}
\end{center}}
\end{table}

\newpage

\begin{table}
\caption[ ]{\label{derivedB2.tab}Derived properties of the cluster sample
in M82 B2}
{\scriptsize
\begin{center}
\begin{tabular}{rrrrrrrr}
\hline
\hline
\multicolumn{1}{c}{\#} & \multicolumn{1}{c}{$A_V$} &
\multicolumn{1}{c}{$M_V^0$} & \multicolumn{1}{c}{$(B-V)_0$} &
\multicolumn{1}{c}{$(V-I)_0$} & \multicolumn{1}{c}{$(V-H)_0$} &
\multicolumn{1}{c}{$(J-H)_0$} & \multicolumn{1}{c}{log( Age )} \\
\cline{2-7}
& \multicolumn{6}{c}{(mag)} & \multicolumn{1}{c}{(yr)} \\
\hline
 1 & 0.00 & $ -5.79$ & $ 1.06$ & $ 1.15$ & $ 2.48$ & $  0.06$ & $10.0$ \\
 2 &$\dots$&$<-7.57$ & $\dots$ & $<3.02$ & $<6.49$ & $ <1.01$ &$\dots$ \\
 3 &$\dots$&$<-6.43$ & $\dots$ & $<2.29$ & $\dots$ & $ \dots$ &$\dots$ \\
 4 & 0.46 & $ -8.14$ & $ 0.74$ & $ 0.99$ & $<3.98$ & $ <1.12$ & $ 9.3$ \\
 5 & 0.99 & $ -9.02$ & $ 0.32$ & $ 0.65$ & $ 0.97$ & $  0.06$ & $ 8.7$ \\
 6 &$\dots$&$<-6.47$ & $\dots$ & $<1.85$ & $<3.94$ & $ \dots$ &$\dots$ \\
 7 & 1.34 & $ -8.56$ & $ 0.72$ & $ 1.05$ & $<0.43$ & $ \dots$ & $ 9.2$ \\
 8 & 2.39 & $ -8.19$ & $ 0.92$ & $ 1.32$ & $<2.82$ & $ <0.37$ & $ 9.7$ \\
 9 &$\dots$&$<-5.49$ & $<0.49$ & $<1.78$ & $<4.61$ & $ <2.62$ & $<8.8$ \\
10 & 1.49 & $ -7.19$ & $>0.65$ & $ 1.00$ & $ 2.51$ & $  0.39$ & $ 9.4$ \\
11 & 2.38 & $ -7.72$ & $>0.25$ & $ 0.73$ & $ 1.81$ & $  0.32$ & $ 8.9$ \\
12 & 0.69 & $-10.59$ & $ 0.92$ & $ 1.14$ & $ 2.75$ & $  0.52$ & $ 9.7$ \\
13 & 0.58 & $ -7.54$ & $ 0.93$ & $ 1.16$ & $ 2.36$ & $  0.35$ & $ 9.8$ \\
14 & 1.33 & $ -6.68$ & $ 0.89$ & $ 1.18$ & $<4.59$ & $ <1.40$ & $10.1$ \\
15 & 0.58 & $ -7.59$ & $ 1.37$ & $ 2.11$ & $ 4.23$ & $  0.74$ &$\dots$ \\
16 & 3.40 & $ -8.88$ & $>0.02$ & $ 0.85$ & $<2.40$ & $ <0.43$ & $ 9.0$ \\
17 & 0.81 & $ -7.85$ & $ 0.86$ & $ 1.12$ & $ 2.53$ & $  0.55$ & $ 9.5$ \\
18 & 3.18 & $ -8.81$ & $ 0.32$ & $ 0.94$ & $<2.25$ & $ <0.68$ & $ 9.2$ \\
19 &$\dots$&$<-7.12$ & $\dots$ & $<2.88$ & $<6.09$ & $ <0.98$ &$\dots$ \\
20 &$\dots$&$<-6.28$ & $\dots$ & $<1.72$ & $<3.02$ & $ <0.34$ &$\dots$ \\
21 & 3.49 & $ -9.99$ & $ 0.12$ & $ 0.76$ & $<2.83$ & $ <0.59$ & $ 8.9$ \\
22 & 0.48 & $ -5.86$ & $ 1.06$ & $ 1.26$ & $<3.01$ & $ <1.79$ & $10.3$ \\
23 & 1.22 & $ -6.56$ & $ 0.76$ & $ 1.07$ & $ 2.61$ & $  0.57$ & $ 9.7$ \\
24 & 1.87 & $ -8.43$ & $>0.42$ & $ 0.86$ & $ 2.07$ & $  0.31$ & $ 9.0$ \\
25 & 0.89 & $ -8.37$ & $ 0.36$ & $ 0.69$ & $ 2.20$ & $  0.61$ & $ 8.7$ \\
26 & 0.42 & $ -9.65$ & $ 0.40$ & $ 0.67$ & $ 1.67$ & $  0.37$ & $ 8.8$ \\
27 &$\dots$&$<-5.36$ & $<0.33$ & $<1.24$ & $<3.23$ & $ <0.73$ & $<8.8$ \\
28 & 1.85 & $ -7.87$ & $ 0.05$ & $ 0.53$ & $ 1.51$ & $  0.46$ & $ 7.3$ \\
29 & 0.78 & $ -6.27$ & $ 0.26$ & $ 0.58$ & $ 1.21$ & $  0.56$ & $ 8.7$ \\
30 & 1.14 & $ -8.07$ & $ 0.53$ & $ 0.85$ & $ 1.88$ & $  0.33$ & $ 8.9$ \\
31 & 0.75 & $ -7.05$ & $ 0.33$ & $ 0.65$ & $<2.30$ & $ <1.46$ & $ 8.7$ \\
32 & 1.16 & $ -7.48$ & $ 0.79$ & $ 1.10$ & $ 2.28$ & $  0.27$ & $ 9.8$ \\
33 & 0.75 & $ -6.08$ & $ 0.48$ & $ 0.79$ & $<3.53$ & $ <1.17$ & $ 8.8$ \\
34 & 0.89 & $ -6.21$ & $ 0.27$ & $ 0.62$ & $ 2.08$ & $  0.54$ & $ 8.6$ \\
35 & 1.91 & $ -7.67$ & $>0.20$ & $ 0.67$ & $ 2.18$ & $  0.54$ & $ 8.8$ \\
36 & 0.00 & $ -8.08$ & $ 0.73$ & $ 0.90$ & $ 2.20$ & $  0.47$ & $ 9.2$ \\
37 & 2.14 & $-10.13$ & $ 0.05$ & $ 0.55$ & $ 1.92$ & $  0.37$ & $ 7.5$ \\
38 & 0.00 & $ -7.14$ & $ 0.45$ & $ 0.57$ & $ 1.80$ & $  0.69$ & $ 8.8$ \\
39 & 0.00 & $ -6.57$ & $ 0.46$ & $ 0.57$ & $<3.16$ & $ <1.51$ & $ 8.8$ \\
40 & 0.39 & $ -8.88$ & $ 0.26$ & $ 0.55$ & $ 1.54$ & $  0.39$ & $ 8.5$ \\
41 & 0.62 & $ -9.86$ & $ 0.63$ & $ 0.86$ & $ 2.17$ & $  0.24$ & $ 9.0$ \\
42 &$\dots$&$<-6.32$ & $<1.91$ & $<2.30$ & $<4.47$ & $ <0.97$ &$\dots$ \\
43 & 0.00 & $ -6.89$ & $ 0.44$ & $ 0.57$ & $<2.17$ & $ <1.71$ & $ 8.8$ \\
44 & 0.37 & $ -7.26$ & $ 0.35$ & $ 0.63$ & $<2.24$ & $ <1.98$ & $ 8.7$ \\
45 & 1.16 & $ -7.28$ & $ 0.79$ & $ 1.13$ & $ 1.97$ & $  0.40$ & $ 9.9$ \\
46 & 0.21 & $ -5.84$ & $ 0.06$ & $<0.51$ & $<2.57$ & $ \dots$ & $ 7.5$ \\
47 & 0.00 & $ -5.81$ & $ 0.76$ & $ 0.93$ & $<1.02$ & $<-0.05$ & $ 9.3$ \\
48 & 0.73 & $ -6.69$ & $ 0.31$ & $ 0.63$ & $ 1.86$ & $  0.58$ & $ 8.7$ \\
49 & 0.46 & $ -8.30$ & $ 0.86$ & $ 1.07$ & $ 2.18$ & $  0.21$ & $ 9.5$ \\
50 & 1.49 & $ -7.83$ & $ 0.15$ & $ 0.57$ & $ 1.62$ & $  0.37$ & $ 8.3$ \\
\multicolumn{4}{l}{(continued on next page)} \\
\hline
\end{tabular}
\end{center}}
\end{table}

\newpage

\addtocounter{table}{-1}
\begin{table}
\caption[ ]{(continued)}
{\scriptsize
\begin{center}
\begin{tabular}{rrrrrrrr}
\hline
\hline
\multicolumn{1}{c}{\#} & \multicolumn{1}{c}{$A_V$} &
\multicolumn{1}{c}{$M_V^0$} & \multicolumn{1}{c}{$(B-V)_0$} &
\multicolumn{1}{c}{$(V-I)_0$} & \multicolumn{1}{c}{$(V-H)_0$} &
\multicolumn{1}{c}{$(J-H)_0$} & \multicolumn{1}{c}{log( Age )} \\
\cline{2-7}
& \multicolumn{6}{c}{(mag)} & \multicolumn{1}{c}{(yr)} \\
\hline
\multicolumn{4}{l}{(continued from previous page)} \\
51 & 0.00 & $ -5.63$ & $ 0.49$ & $ 0.53$ & $<2.89$ & $ \dots$ & $ 8.6$ \\
52 & 0.00 & $ -6.87$ & $ 0.65$ & $ 0.77$ & $ 0.84$ & $ -0.10$ & $ 9.0$ \\
53 &$\dots$&$<-7.42$ & $<0.80$ & $<0.78$ & $<1.86$ & $ <0.56$ & $<9.0$ \\
54 & 0.00 & $ -6.89$ & $ 0.33$ & $ 0.57$ & $ 1.54$ & $  0.36$ & $ 8.7$ \\
55 & 0.00 & $ -5.65$ & $ 0.42$ & $ 0.51$ & $<2.57$ & $ <1.03$ & $ 8.8$ \\
56 & 0.00 & $ -7.03$ & $ 0.75$ & $ 0.96$ & $ 2.91$ & $  0.83$ & $ 9.2$ \\
57 & 0.81 & $ -8.02$ & $ 0.76$ & $ 1.02$ & $ 2.85$ & $  0.33$ & $ 9.3$ \\
58 &$\dots$&$<-7.99$ & $\dots$ & $<1.67$ & $<3.14$ & $ <0.46$ &$\dots$ \\
59 & 2.03 & $ -7.68$ & $ 0.38$ & $ 0.83$ & $ 1.70$ & $  0.33$ & $ 8.8$ \\
60 & 0.57 & $ -6.81$ & $ 0.41$ & $ 0.68$ & $ 1.74$ & $  0.59$ & $ 8.8$ \\
61 & 0.42 & $ -6.25$ & $ 0.34$ & $<0.62$ & $ 1.00$ & $  0.13$ & $ 8.7$ \\
62 & 0.00 & $ -5.50$ & $ 0.58$ & $ 0.76$ & $ 1.24$ & $  0.04$ & $ 9.0$ \\
63 & 0.52 & $ -7.39$ & $ 0.37$ & $ 0.66$ & $ 1.61$ & $  0.29$ & $ 8.8$ \\
64 & 0.00 & $ -5.44$ & $ 0.78$ & $ 0.94$ & $ 2.46$ & $  0.97$ & $ 9.3$ \\
65 & 0.62 & $ -7.72$ & $ 0.67$ & $ 0.92$ & $ 1.47$ & $  0.25$ & $ 9.0$ \\
66 &$\dots$&$<-5.14$ & $<0.52$ & $<0.77$ & $<2.72$ & $ <0.83$ & $<9.0$ \\
67 & 1.08 & $ -8.24$ & $ 0.57$ & $ 0.87$ & $ 2.37$ & $  0.56$ & $ 8.9$ \\
68 & 0.21 & $ -6.38$ & $ 0.73$ & $ 0.96$ & $ 2.43$ & $  0.53$ & $ 9.2$ \\
69 & 0.00 & $ -5.61$ & $ 0.57$ & $ 0.70$ & $ 1.96$ & $  0.48$ & $ 8.9$ \\
70 & 0.10 & $ -7.97$ & $ 0.44$ & $ 0.68$ & $<1.76$ & $ <0.81$ & $ 8.8$ \\
\hline
\end{tabular}
\end{center}}
\end{table}

\newpage

\begin{table}
\caption[ ]{\label{colpeaks.tab}Color distributions of the cluster
candidates in M82, the globular cluster population in the Milky Way and
young star cluster populations in comparison galaxies}
{\scriptsize
\begin{center}
\begin{tabular}{lllccll}
\hline
\hline
\multicolumn{1}{c}{Galaxy} & \multicolumn{1}{c}{Sample} &
\multicolumn{1}{c}{Color$^1$} & \multicolumn{1}{c}{Mean} &
\multicolumn{1}{c}{Dispersion} & \multicolumn{1}{c}{Age
range} & \multicolumn{1}{c}{References$^2$} \\
& & & \multicolumn{1}{c}{(mag)} & $\sigma$ (mag) &
\multicolumn{1}{c}{(Gyr, for $Z_\odot$)} \\
\hline
M82       & B1                    & $(B-V)_0$     & 0.78 & 0.29 & $0.35-1.4$\\
          &                       & $(V-I)_0$     & 1.13 & 0.40 \\
M82       & B2                    & $(B-V)_0$     & 0.87 & 0.45 & $0.35-1.4$\\
          &                       & $(V-I)_0$     & 1.33 & 0.62 \\
\hline
\multicolumn{7}{c}{\underline{Local, well-studied comparison samples}} \\
Milky Way & Total                 & $(B-V)_{0,c}$ & 0.71 & 0.12 & 12 -- 14  & R97,R98,vdB92 \\
          &                       & $(V-I)_{0,c}$ & 0.84 & 0.13 \\
LMC       & Total                 & $(B-V)_0$     & 0.20 & 0.30 & 0.03 -- 10 & EF88  \\
          &                       &               &      &      & (mean $\sim 0.3$) \\
\hline
\multicolumn{7}{c}{\underline{Young star cluster populations based on {\sl HST} photometry}}\\
NGC 1275  & $^3$                  & $(B-V)_0$     & 0.03 & 0.32 & $\lesssim 0.3$ & H92, R93  \\
          &                       & $(V-I)_0$     & 0.09 & 0.55 & (mean $\sim 10^7$ yr)    \\
NGC 3921  & $\sigma_{V-I}\le0.25$ & $(V-I)_0$     & 0.69 &      & 0.08 -- 0.50 & S96  \\
          & Associations          & $(V-I)_0$     & 0.59 &      & (median $\sim 0.25$) & S96  \\
NGC 4038/39 & Young clusters      & $(B-V)_0$     & 0.17 & 0.22 & $< 0.030$   & W99 \\
          & Young clusters        & $(V-I)_0$     & 0.38 & 0.30 & $< 0.030$   & W99 \\
          & Interm.-age cl.       & $(B-V)_0$     & 0.30 & 0.27 & 0.25 -- 1.0 & W99 \\
          & Interm.-age cl.       & $(V-I)_0$     & 0.53 & 0.35 & 0.25 -- 1.0 & W99 \\
NGC 7252  & $V_0 < 24$            & $(B-V)_{0,1}$ & 0.65 &      & 0.005 -- 0.040 & W93, M98 \\
          &                       & $(V-I)_{0,1}$ & 0.4  &      & (mean $\sim 0.034$) & W93, M98 \\
          &                       & $(B-V)_{0,2}$ & 1.0  &      & 0.50 -- 0.80 & W93, M98 \\
          &                       & $(V-I)_{0,2}$ & 0.8  & 0.25 & (mean 0.65) & W93, M98 \\
\hline
\end{tabular}
\end{center}
}
$^1$ Correction for foreground extinction is indicated by the subscript
0; internal extinction correction by the subscript 'c' \\
$^2$ This paper if no reference given; EF88: Elson \& Fall (1988); H92:
Holtzman et al. (1992); M98: Miller et al. (1998); R93: Richer et al.
(1993); R97, R98: Reid (1997, 1998); S96: Schweizer et al. (1996);
vdB92: vandenBergh (1992); W93: Whitmore et al. (1993); W99: Whitmore et
al. (1999) \\
$^3$ Cross-correlated sources with those of Holtzman et al. (1992) \\
\end{table}

\begin{table}
\caption[ ]{\label{bright.tab}Characteristics of the brightest star
cluster candidates}
\begin{center}
\begin{tabular}{ccrcccc}
\hline
\hline
Region & Cluster & \#$^1$ & \multicolumn{1}{c}{$V_0$} &
\multicolumn{1}{c}{FWHM} & \multicolumn{2}{c}{$R_{\rm core}$ (pc)} \\
\cline{6-7}
& & & \multicolumn{1}{c}{(mag)} & \multicolumn{1}{c}{(pc)} & $\gamma =
2.0$ & $\gamma = 2.6$ \\
\hline
B1 & (1) & 28 & $17.84 \pm 0.01$ & 4.3 & 1.7 & 2.1 \\
   & (2) & 18 & $18.92 \pm 0.02$ & 4.3 & 1.7 & 2.2 \\
   & (3) & 11 & $19.06 \pm 0.03$ & 4.3 & 1.4 & 1.8 \\
   & (4) &  7 & $19.08 \pm 0.03$ & 3.5 & 1.6 & 1.8 \\
   & (5) &  8 & $19.10 \pm 0.02$ & 5.4 & 2.0 & 2.6 \\
B2 & (1) & 12 & $17.81 \pm 0.01$ & 4.2 & 2.0 & 2.2 \\
   & (2) & 41 & $18.47 \pm 0.02$ & 4.3 & 1.5 & 2.0 \\
   & (3) & 26 & $18.48 \pm 0.01$ & 5.4 & 2.0 & 2.5 \\
   & (4) & 36 & $19.72 \pm 0.01$ & 7.9 & 2.9 & 3.3 \\
   & (5) & 40 & $19.22 \pm 0.01$ & 5.4 & 2.2 & 2.6 \\
\hline
& & & \multicolumn{1}{r}{\it mean:} & $4.9 \pm 1.2$ & $1.9 \pm 0.4$ &
$2.3 \pm 0.4$ \\
\hline
\end{tabular}
\end{center}
$^1$ -- See Tables 4 and 5. 
\end{table}

\end{document}